\documentclass[aps,prd,nofootinbib,showpacs,preprintnumbers]{revtex4}
\usepackage[usenames]{color}
\usepackage[dvips]{epsfig}
\usepackage{amssymb}
\usepackage{verbatim}
\usepackage{psfrag}
\newcommand{\beqn} {\begin{equation}}
\newcommand{\eqn} {\end{equation}}
\newcommand{\hm}{\hat{m}}

\newcommand{\expval}[1]{\langle #1\rangle}
\def \beq{\begin{equation}}
\def \eeq{\end{equation}}
\def \bea{\begin{eqnarray}}
\def \eea{\end{eqnarray}}

\def \bet0{\beta_0}
\def \bet1{\beta_1}
\def \simgt{\,\rlap{\lower 7.5 pt\hbox{$\mathchar \sim$}}\raise 3 pt \hbox{$>$}\,}
\def \simlt{\,\rlap{\lower 7.5 pt\hbox{$\mathchar \sim$}}\raise 3 pt \hbox{$<$}\,}

\def\lsim{\raise0.3ex\hbox{$<$\kern-0.75em\raise-1.1ex\hbox{$\sim$}}}
\def\gsim{\raise0.3ex\hbox{$>$\kern-0.75em\raise-1.1ex\hbox{$\sim$}}}

\begin{document}
%\linenumbers
\title{On the magnetic equation of state in (2+1)-flavor QCD}
\author{
S. Ejiri$^{\rm a}$, F. Karsch$^{\rm a,b}$, 
E. Laermann$^{\rm b}$, C. Miao$^{\rm a}$, S. Mukherjee$^{\rm a}$, \\
P. Petreczky$^{\rm a,c}$, C. Schmidt$^{\rm b}$, W. Soeldner$^{\rm d}$, W. Unger$^{\rm b}$
}
\affiliation{
$^{\rm a}$ Physics Department, Brookhaven National Laboratory,Upton, NY 11973, USA \\
$^{\rm b}$ Fakult\"at f\"ur Physik, Universit\"at Bielefeld, D-33615 Bielefeld, Germany\\
$^{\rm c}$ RIKEN-BNL Research Center, Brookhaven National Laboratory, Upton, NY 11973, USA \\
$^{\rm d}$ ExtreMe Matter Institute EMMI, GSI Helmholtzzentrum f\"ur 
Schwerionenforschung, Planckstr.~1, D-64291 Darmstadt, Germany \\
}

\begin{abstract}
A first study of critical behavior in the vicinity of
the chiral phase transition of (2+1)-flavor QCD is presented.  
We analyze the quark mass and volume dependence of the chiral condensate 
and chiral susceptibilities in QCD with two degenerate light quark masses 
and a strange quark. The strange quark mass ($m_s$) is chosen close to its 
physical value; the two degenerate light quark masses ($m_l$) are varied in 
a wide range $1/80 \le m_l/m_s \le 2/5$, where the smallest light quark
mass value corresponds to a pseudo-scalar Goldstone mass of about 75~MeV.
All calculations are performed with staggered fermions on lattices with 
temporal extent $N_\tau=4$.
We show that numerical results are consistent with $O(N)$ scaling in the
chiral limit. We find that in the region of physical light quark mass values, 
$m_l/m_s \simeq 1/20$, the temperature and quark mass dependence of the
chiral condensate is already dominated by universal properties of QCD that are encoded
in the scaling function for the chiral order parameter, {\it the magnetic
equation of state}.  
We also provide evidence for the influence of thermal fluctuations of Goldstone
modes on the chiral condensate at finite temperature. 
At temperatures below, but close to the chiral phase transition at vanishing
quark mass, this leads to a characteristic dependence of the light quark 
chiral condensate on the square root of the light quark mass.
\end{abstract}
\maketitle

\section{Introduction}

Chiral symmetry and its spontaneous breaking in the vacuum
are key ingredients to our understanding of the phase 
structure of strongly interacting matter at non-zero temperature
and vanishing baryon chemical potential. In the limit of $n_f$
massless quark flavors the QCD phase transition is 
controlled by the $SU_L(n_f)\times SU_R(n_f)$ chiral
symmetry. Quite general renormalization group arguments 
suggest \cite{Pisarski} that QCD with three degenerate light 
quark flavors has a first order phase transition,
whereas the 2-flavor theory is expected to have a second order 
phase transition. In the latter case the  $SU_L(2)\times SU_R(2)$ 
chiral symmetry is isomorphic to $O(4)$ and the transition therefore
is expected to belong to the same universality class as 3-dimensional,
$O(4)$ symmetric spin models. Depending on the value of the strange
quark mass the QCD phase transition in the limit of vanishing
light quark masses (up, down) may be first order or a continuous
transition still belonging to the  3-dimensional, $O(4)$ universality
class \cite{Pisarski}. 

While numerical calculations in 3-flavor QCD gave evidence for the
existence of a first order transition,
many of the details of the transition in 2- or (2+1)-flavor QCD 
with light up and down quarks are still poorly constrained through lattice
calculations. In particular, we do not know whether the chiral phase transition
in (2+1)-flavor QCD is first or second order. An answer to this 
question is not only of academic interest; it also greatly influences 
our thinking about the phase diagram of QCD at non-zero baryon chemical potential \cite{Gupta}.
The present analysis, although still performed on rather coarse lattices,
is a first step towards answering this question.

Attempts to verify the universal critical behavior associated with the
QCD chiral phase transition in 2-flavor QCD have been made already 
before in calculations with 
staggered \cite{Karsch94,KL,Bernard,Aoki,Bernard00,DiGiacomo,Mendes} and
Wilson \cite{Iwasaki,AliKhan} fermions.  
None of these lattice discretization schemes for the fermion sector of QCD
preserve the full chiral $O(4)$ symmetry of the QCD Lagrangian. 
It therefore may not be too surprising that the early attempts to verify
universal scaling properties of QCD were not too successful. 
In fact,
at non-zero lattice spacing the Wilson fermion formulation does not preserve 
any continuous symmetry related to the chiral sector of QCD. The staggered
formulation preserves at least an $O(2)$ symmetry 
at non-zero lattice spacing
that gives rise to a single massless Goldstone mode in the chiral limit.
Nevertheless,
direct determinations of critical exponents within the staggered discretization
scheme \cite{Karsch94,KL,Aoki} did not  
deliver the expected $O(N)$ results. Of course, due to the explicit breaking of $O(4)$
symmetry in the staggered formalism at non-vanishing lattice spacing one would not 
have expected to be sensitive to $O(4)$ scaling. However, $O(4)$ and $O(2)$ critical 
exponents are quite similar and one thus might have hoped to observe at least some
{\it generic evidence for $O(N)$ scaling}\footnote{We are dealing here with 
numerical calculations of a cutoff theory whose Lagrangian
has a global $O(2)$ symmetry. The relevant symmetry of QCD in the
continuum limit, on the other hand, is expected to be
$O(4)$. For many aspects of the 
discussion of critical behavior presented here, the distinction between $O(2)$
and $O(4)$ is of no importance. In these cases we will generically talk about
$O(N)$ symmetric models.}. 

In the same spirit, the magnetic 
equation of state, {\it i.e.} the scaling of the chiral order parameter as function
of reduced temperature and quark mass, has been analyzed subsequently in calculations
with staggered \cite{Bernard,Bernard00,DiGiacomo} 
and Wilson fermions \cite{Iwasaki,AliKhan}.
The studies performed with Wilson fermions gave some indication for $O(4)$ 
scaling. These calculations, however, had been constrained
to the high temperature, symmetry restored phase and had been performed with 
rather large values of the quark mass. They therefore did not allow to perform
a test of scaling in the symmetry broken phase and could not contribute
to the 
question of how Goldstone modes influence the scaling behavior 
at low temperatures.
This contribution of Goldstone modes is a 
prominent feature of the magnetic equation of state of $O(N)$ symmetric theories 
which
has been analyzed in detail in $O(N)$ symmetric spin models 
\cite{Toussaint,EngelsO2,EngelsO4,Engels2001,Engels2003}.
In the case of staggered fermions one might have
hoped to find at least evidence for 
$O(2)$ scaling\footnote{Convincing evidence for $O(2)$ scaling
has been found in calculations with a massless staggered fermion action to 
which an irrelevant chiral 4-fermion interaction has been added \cite{Kogut}.}. 
In fact, not only critical
exponents, but also the $O(2)$ and $O(4)$ magnetic equations of state are quite 
similar. Deviations from the scaling function, however, turned out to be large 
in the low as well as high temperature regions and even in calculations on lattices 
with rather small lattice spacings \cite{Bernard,Bernard00}. 
The missing evidence for $O(N)$ scaling
also left room for an interpretation of the scaling behavior of 2-flavor
QCD in terms of a first order phase transition \cite{DiGiacomo}.

We will present here results from calculations with staggered fermions.
Contrary to most earlier studies of scaling properties these 
calculations
have been performed with an action that suppresses cut-off effects induced
by a non-zero lattice spacing ($a$) in finite temperature calculations.
Thermodynamic quantities are $O(a^2)$ improved. 
A first analysis of Goldstone effects with this action
has been performed for rather large quark masses in Ref.~\cite{Engels_Gold}.
Our calculations have been carried out with smaller than physical light quark masses
so that the lightest Goldstone mode is a factor two lighter than the
physical value of the pion mass. This will allow us to address
the basic features of the thermodynamics induced by Goldstone modes and 
to gain some
control over the universal features in the vicinity of the chiral 
phase transition temperature. 
We will present numerical evidence that
at finite temperature, in the symmetry broken phase, the dominant 
quark mass dependence of the 
chiral condensate arises from fluctuations
of the Goldstone modes that lead to a square root dependence of the condensate
on the light quark masses. Such a behavior is expected in 3-dimensional theories 
with global $O(N)$ symmetry 
\cite{Wallace,Leutwyler,Hasenfratz,Smilga}.
We will also show that universal scaling properties of the condensate
are consistent with the 3-dimensional $O(N)$ universality class. Furthermore,
an analysis of scaling violations, induced by the regular part of the QCD 
partition function, suggests that the crossover transition in QCD with 
physical quark masses is already strongly influenced by contributions arising
from the singular universal part of the QCD partition function.  
At present we
are not sensitive to differences between $O(2)$ and $O(4)$ scaling. However, we 
point out that a combined analysis of scaling functions for the order parameter
and its susceptibility should provide unambiguous results on the universality
class of the chiral transition in QCD.

This paper is organized as follows. In the next section we summarize 
universal properties of 3-dimensional $O(2)$ and $O(4)$ symmetric spin models
and introduce notations. In Section III we present our data
on the quark mass and temperature dependence of chiral condensates in 
(2+1)-flavor QCD. 
The main 
results on the magnetic equation of state
are discussed in Section IV. In Section V we give a brief 
account of properties
of susceptibilities of the chiral order parameter. Section VI contains our 
conclusions. 
In Appendix \ref{app.scaling},
for the readers' convenience
we compile the asymptotic forms and the interpolations used for the
$O(2)$ and $O(4)$ scaling functions, as adopted from 
\cite{Engels2001,Engels2003}.
The numerical data which this paper is based on
are summarized in Appendix \ref{app.data}.

\section{\boldmath$O(N)$ symmetry breaking}

In the limit of vanishing light quark masses QCD is expected to undergo a phase
transition at some critical temperature $T_c$ at which chiral symmetry gets restored.
The light quark chiral condensate, $\langle \bar\psi \psi \rangle_l$, will 
vanish at this temperature. Its quark mass and temperature dependence in the vicinity
of the critical point, $(T,m_l)\equiv (T_c,0)$, is controlled by a scaling function
that arises from the singular part of the partition function. 
One way of analyzing the non-analytic structure of the QCD partition function, which
has been pursued in the past, is to study the so-called magnetic equation of state.
Before continuing our 
discussion of critical behavior in QCD we briefly summarize basic scaling relations
using the conventional spin model notation, where the order parameter is denoted
by $M$ and the symmetry breaking field is denoted by $H$. The critical 
behavior of $O(2)$ and $O(4)$ spin models in three dimensions has been analyzed extensively 
in the past. We will follow here closely the discussion given in \cite{Engels2001}.

\subsection{Magnetic Equation of State}

In the vicinity of a critical point regular contributions to the partition
functions become negligible and 
the universal critical behavior of the order parameter $M$ of, e.g. 3-dimensional
$O(N)$ spin models, is controlled by a scaling function $f_G$ that arises from
the singular part of the logarithm of the partition function, 
\begin{equation}
M(t,h) \;=\; h^{1/\delta} f_G(z) \; ,
\label{order}
\end{equation} 
with  $z=t/h^{1/\beta\delta}$ and scaling variables $t$ and $h$ that are related to 
the temperature, $T$, and the symmetry breaking (magnetic) field, $H$,  
\begin{equation}
t = \frac{1}{t_0}\frac{T-T_c}{T_c} \quad , \quad h= \frac{H}{h_0} \; .
\label{reduced}
\end{equation}
Here $\beta$ and $\delta$ are critical exponents characterizing
the approach of the order parameter $M$ to the critical point when 
one of the scaling variables is set to zero,
\begin{eqnarray}
M&=& (-t)^\beta \quad ,\quad h\equiv 0  \; ,\;  t<0\label{tscaling}\\
M&=& h^{1/\delta}  \quad ,\quad\;\; t \equiv 0 \label{hscaling}
\end{eqnarray}
These relations also fix the normalization of the scaling variables $t$ and $h$, 
{\it i.e.} they define the normalization constants $t_0$ and $h_0$ introduced in 
Eq.~\ref{reduced}. 
Equivalently one can fix $t_0$ and $h_0$ through normalization
conditions for the scaling function $f_G$,
\begin{equation}
f_G(0) = 1 \quad , \quad \lim_{z\rightarrow -\infty} \frac{f_G(z)}{(-z)^\beta} = 1 \; .
\label{fGscaling}
\end{equation}
As it is of some relevance for our later discussion we note here that the normalization 
conditions for the scaling function, $f_G$, which fix the scale parameters $t_0$ and $h_0$, refer 
to values of the scaling variable $z$, that are infinitely apart. 

\begin{figure}[t]
\begin{center}
\epsfig{file=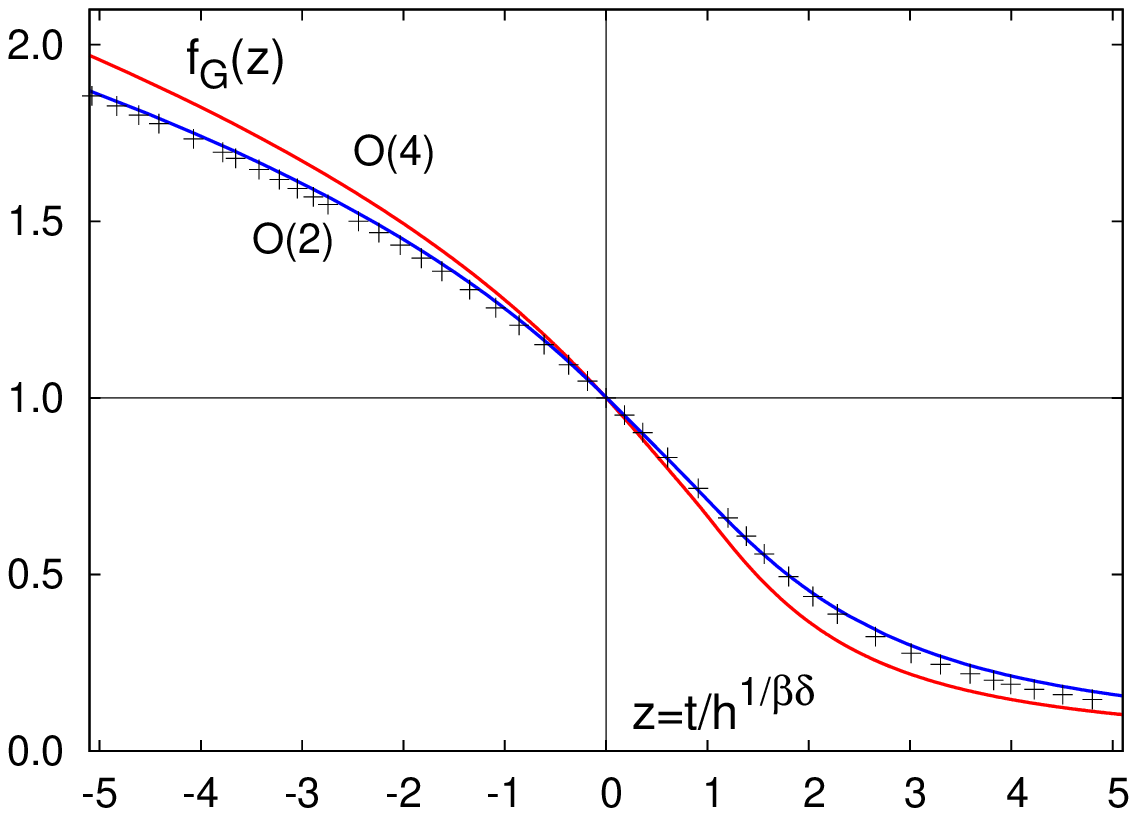,width=9.0cm}\hspace*{-0.4cm}\epsfig{file=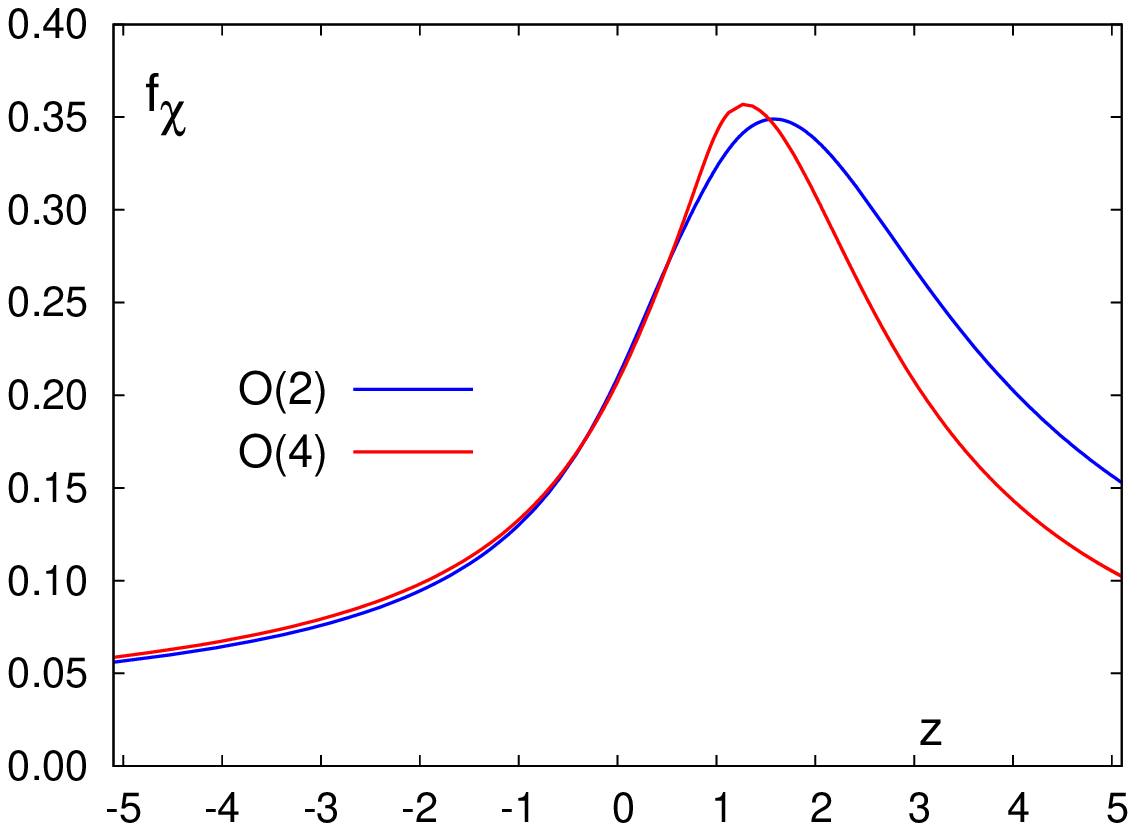,width=9.0cm}
\end{center}
\caption{\label{fig:scaling} Scaling functions for the universality classes
of three dimensional $O(2)$ and $O(4)$ models (left). 
Crosses show the $O(4)$ scaling 
function with an argument $\tilde{z}=1.2 z$. 
The right hand figure shows the scaling
function for the chiral susceptibility introduced in Eq.~\protect\ref{fchi}.
}
\end{figure}
 
For $O(N)$ symmetric spin models in three dimensions the scaling functions have been
analyzed in much detail using Monte Carlo simulations and renormalization group 
techniques. In Fig.~\ref{fig:scaling} we show results for the 
$O(2)$ ~\cite{EngelsO2} and $O(4)$ \cite{Toussaint,EngelsO4}
scaling functions obtained by using the implicit parametrizations given in 
Ref.~\cite{Engels2001} and compiled in Appendix~\ref{app.scaling}.
There are clear differences between both scaling
functions. We note, however, that the manifestation of the differences between 
the $O(2)$ and $O(4)$ scaling functions in the limited range $|z|\le 5$
shown in Fig.~\ref{fig:scaling} relies also on the normalization of these 
scaling functions in the limit $z\rightarrow -\infty$. Without information on the
scaling function at $z \rightarrow -\infty$, {\it i.e.} in a numerical study that
only has access to a limited range of $z$-values, 
the scale parameters $t_0$ and $h_0$ could easily be adjusted
to make the $O(2)$ and $O(4)$ scaling functions almost coincide. For 
$|z|\le 5$ this is
shown by the crosses in Fig.~\ref{fig:scaling} which have been obtained by
rescaling the argument of the $O(4)$ scaling function, $z \rightarrow 1.2 z$, {\it i.e.} 
through a change of $z_0$ by 20\% in this interval. This 
will lead to a violation of the normalization condition at $z=-\infty$ by a factor 
$1.2^\beta\simeq 1.07$, for the $O(4)$ scaling function. 

The above discussion makes it evident 
that a numerical analysis of the magnetic equation
of state alone, in a scaling regime as large as $|z|\le 5$, will not allow to 
distinguish $O(2)$ scaling from $O(4)$ scaling, unless the numerical accuracy is 
extraordinary high. Additional information will be needed to distinguish $O(2)$
from $O(4)$ scaling. This can be achieved through accurate control over  the 
chiral limit, $z=-\infty$, or through the analysis of other scaling functions 
like the scaling function, $f_\chi (z)$, for the susceptibility of the
order parameter,
\begin{eqnarray}
\chi_M(t,h) &=& \frac{\partial M}{\partial H} = \frac{1}{h_0}
h^{1/\delta -1} f_\chi(z) \; , \label{chiM}\\
f_\chi(z) &=& \frac{1}{\delta}\left( f_G(z) -\frac{z}{\beta} f'_G (z)\right) \; .
\label{fchi}
\end{eqnarray}
This scaling function is shown in Fig.~\ref{fig:scaling}(right). It has a maximum
at $z_p$ which together with some critical exponents of the $O(N)$ models is
given in Table~\ref{tab:parameter}. It is obvious from
this figure that a simple rescaling of the scale parameter $z$ cannot transform
an $O(2)$ scaling function into that for $O(4)$.

While different $O(N)$ symmetric models are characterized by universal scaling
functions, the scaling variables have to be normalized properly.
The scale parameters
$t_0$ and $h_0$ are not universal. They depend on the $O(N)$ symmetric model under
consideration, the definition of the scaling variables
and also on the absolute normalization of the order parameter $M$.
For instance, a rescaling of the order parameter by a constant factor, 
$M \rightarrow b M$, can be absorbed in a redefinition of the normalization 
constants $t_0 \rightarrow b^{-1/\beta} t_0$ and $h_0\rightarrow b^{-\delta} h_0$. 
This leaves the argument $z$ of the scaling function 
and the  scaling function itself unchanged, $z \rightarrow z$ and
$M/h^{1/\delta} \rightarrow M/h^{1/\delta}$. 
For a given definition of the symmetry breaking field $H$
the combination $z_0 = h_0^{1/\beta\delta}/t_0$
therefore remains unchanged and is unique for any $O(N)$ symmetric model, {\it i.e.}
its value is characteristic for that theory. It, for instance, characterizes the 
$H$-dependence of the pseudo-critical line of transition temperatures, $T_p(H)$, 
determined from the peak position of the order parameter susceptibility,
\begin{equation}
\frac{T_p(H) - T_c}{T_c} = \frac{z_p}{z_0}\; H^{1/\beta\delta} \; .
\label{pseudoH}
\end{equation}
In Section IV
we will determine the corresponding scaling relation for the pseudo-critical
line of (2+1)-flavor QCD from an analysis of the magnetic equation of state.
 
\begin{table}[t]
\begin{center}
\vspace{0.3cm}
\begin{tabular}{|l|l|l|l|l|l|}
\hline
$N$ & $~~\beta$ & $~~~\gamma$ & $~~~\delta$ & $~~~{\widetilde c_2}$ & $z_p$ \\
\hline
2 & 0.349 & 1.319 & 4.780 & 0.592(10) & 1.56(10) \\
4 & 0.380 & 1.453 & 4.824 & 0.666(6) & 1.33(5) \\
\hline
\end{tabular}
\end{center}
\caption{Critical exponents $\beta$, $\gamma$, $\delta$ and the universal constant 
${\widetilde c_2}$
for the three dimensional $O(2)$ universality class are taken from \cite{Engels2001}; 
for $O(4)$ we use the data from \cite{Engels2003}. The three critical exponents are
related through $\gamma = \beta (\delta -1)$. The last column gives the 
location of the maximum of the scaling function $f_\chi$ \cite{Engels2001,Engels2003}.}
\label{tab:parameter}
\end{table}

\subsection{Contribution of Goldstone modes}
The spontaneous breaking of the continuous $O(N)$ symmetry at low temperature gives 
rise to massless Goldstone modes. The fluctuations of these light modes are reflected 
in the non-analytic dependence of the order parameter on the symmetry breaking 
variable, $h$ \cite{Wallace}. 
In three dimensions this leads to \cite{Wallace,Hasenfratz,Smilga} 
\begin{equation}
M(t,h) = M(t,0) + c_2(t) \sqrt{h} +{\cal O} (h) \;\;{\rm for ~all} \; t<0\;.
\label{goldstone}
\end{equation}
This leading correction to the temperature dependence of the order parameter, which
arises from a non-vanishing explicit symmetry breaking ($h>0$), is also reflected in 
the leading correction to the scaling function $f_G$.  
For large, negative values of $z$ one has,
\begin{equation}
f_G(z) \simeq f_G^\infty (z) =
(-z)^\beta ( 1 +  {\widetilde c_2}\beta (-z)^{-\beta\delta/2} ) \; ,\; {\rm for}\; 
z\rightarrow -\infty \;.
\label{asymptotic_eos}
\end{equation}
As has been discussed also in \cite{Engels2001}, Eq.~\ref{asymptotic_eos}
can easily be obtained from the magnetic equation of state derived by 
Wallace and Zia \cite{Wallace}. The universal amplitude ${\widetilde c_2}$ is also
given in Table~\ref{tab:parameter}. In the following we will not make use of the 
scaling behavior of $f_G$ in the opposite limit, 
$z\rightarrow +\infty$. We note, however, that in this limit $f_G$ is
controlled by the critical exponent $\gamma=\beta (\delta-1)$, 
$f_G(z) \sim R_\chi z^{-\gamma}$. The universal parameter $R_\chi$
has been determined for three dimensional 
$O(2)$ \cite{VicariO2} and $O(4)$ \cite{VicariO4} universality classes.

\begin{figure}[t]
\begin{center}
\epsfig{file=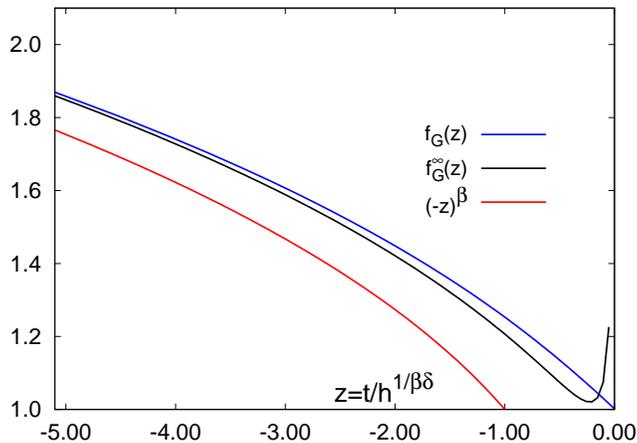,width=9.0cm}
\end{center}
\caption{\label{fig:ON_Goldstone} The O(2) scaling function, $f_G(z)$ compared to the asymptotic
form, $f_G^\infty(z)$ and the leading order term, $(-z)^\beta$. 
}
\end{figure}

The asymptotic form of the $O(N)$ scaling function, $f_G^\infty (z)$, gives an excellent
approximation to $f_G(z)$ in almost the entire low temperature regime, $t<0$. This is 
evident from Fig.~\ref{fig:ON_Goldstone} where we compare $f_G$ to $f_G^\infty$ as well
as to the leading order form, $(-z)^\beta$. For $z <-2$ differences between $f_G$ 
and $f_G^\infty$
are less than 2\% for $O(2)$ and less than 1\% for $O(4)$. The leading 
$h$-dependent correction that arises from the presence of Goldstone modes 
thus gives the dominant contribution to the order parameter in this regime. 
In order to establish universal critical behavior through an 
analysis of the $O(N)$ magnetic equation of state for QCD it therefore will be crucial to
establish the influence of the Goldstone mode on the quark mass dependence of the chiral
condensate and its derivative, the chiral susceptibility.

\section{Chiral symmetry breaking in (2+1) flavor QCD}

\subsection{Quark mass and volume dependence of the chiral condensate}
We discuss here our calculations performed for (2+1)-flavor QCD on lattices
with size $N_\sigma^3\times N_\tau$. We have fixed the temporal extent, $N_\tau =4$,
and performed calculations for different spatial lattice sizes $N_\sigma = 8,\; 16$
and $32$ to control finite volume effects. All our calculations have been performed
with a tree level improved gauge action and an improved staggered fermion action
(p4-action), which eliminates ${\cal O}(a^2)$ discretization errors in 
thermodynamic observables at the tree level.
The value of the bare strange quark mass
in lattice units has been fixed to $\hm_s =0.065$. In earlier calculations
of the equation of state \cite{rbcBIeos} and the transition temperature \cite{Cheng}, 
performed with the same improved gauge and staggered fermion actions, 
it had been shown that 
in the present (small) range of gauge couplings
this value of the strange bare quark mass yields almost physical values 
for the masses of
the strange pseudo-scalar meson and the kaon. The light
quark masses have been varied in the range $0.0008125 \le \hm_l \le 0.026$. The 
smallest value, $m_l/m_s=1/80$, corresponds to a pseudo-scalar Goldstone mass
of about $75$~MeV. For each of the 6 quark mass values chosen in the above interval
we have performed calculations at several values of the gauge coupling in the 
range $3.28\le \beta \le 3.33$. As will become clear later this covers a temperature 
range $0.96 \lsim T/T_c \lsim 1.06 $, with $T_c$ denoting the transition temperature
in the chiral limit on a lattice with fixed temporal extent. All calculations
have been performed using  the Rational Hybrid Monte Carlo (RHMC) algorithm. 
In most cases we collected 15.000 to 40.000 trajectories with length of half a time unit.
We give more details on our simulation parameters, the statistics collected and
expectation values of chiral condensates in Appendix~\ref{app.data}. 

Whenever we convert results to physical units we use the scale setting and meson mass
calculations performed in connection with our calculation of the equation of state
\cite{rbcBIeos} and the transition temperature \cite{Cheng}. The most 
central anchor point for our current analysis is the 
determination of the pseudo-scalar
mass ($m_{ps}$) and the Sommer scale parameter $r_0$ for 
light quark masses $m_l/m_s=1/20$ 
at a gauge coupling $\beta=3.30$. 
This value of the gauge coupling is close to the
critical temperature in the chiral limit and the light to strange quark mass ratio is
close to the physical mass ratio. For this parameter set we find
$m_{ps}a =0.1888(6)$ and $r_0/a= 1.8915(59)$ in lattice units.
This converts to $m_{ps}=150.2(3)$~MeV when using $r_0=0.469$~fm
\cite{Gray} as it has been done also in our earlier calculations. 

The main part of our analysis is based on calculations of the light and strange
quark chiral condensates,
\begin{eqnarray}
\langle \bar\psi \psi \rangle_l &=& \frac{1}{4} \frac{1}{N_\sigma^3 N_\tau} 
\frac{\partial \ln Z}{\partial \hm_l} = \frac{1}{4}
\frac{1}{N_\sigma^3 N_\tau} \langle {\rm Tr} D_l^{-1} \rangle \;\;  , \nonumber \\
\langle \bar\psi \psi \rangle_s &=& \frac{1}{4} \frac{1}{N_\sigma^3 N_\tau} 
\frac{\partial \ln Z}{\partial \hm_s} = \frac{1}{4} 
\frac{1}{N_\sigma^3 N_\tau} \langle {\rm Tr} D_s^{-1} \rangle\; ,
\label{condensate}
\end{eqnarray}
where $D_l$ and $D_s$ denote the fermion matrices for light and strange
quarks, respectively.

\begin{figure}[t]
\begin{center}
\epsfig{file=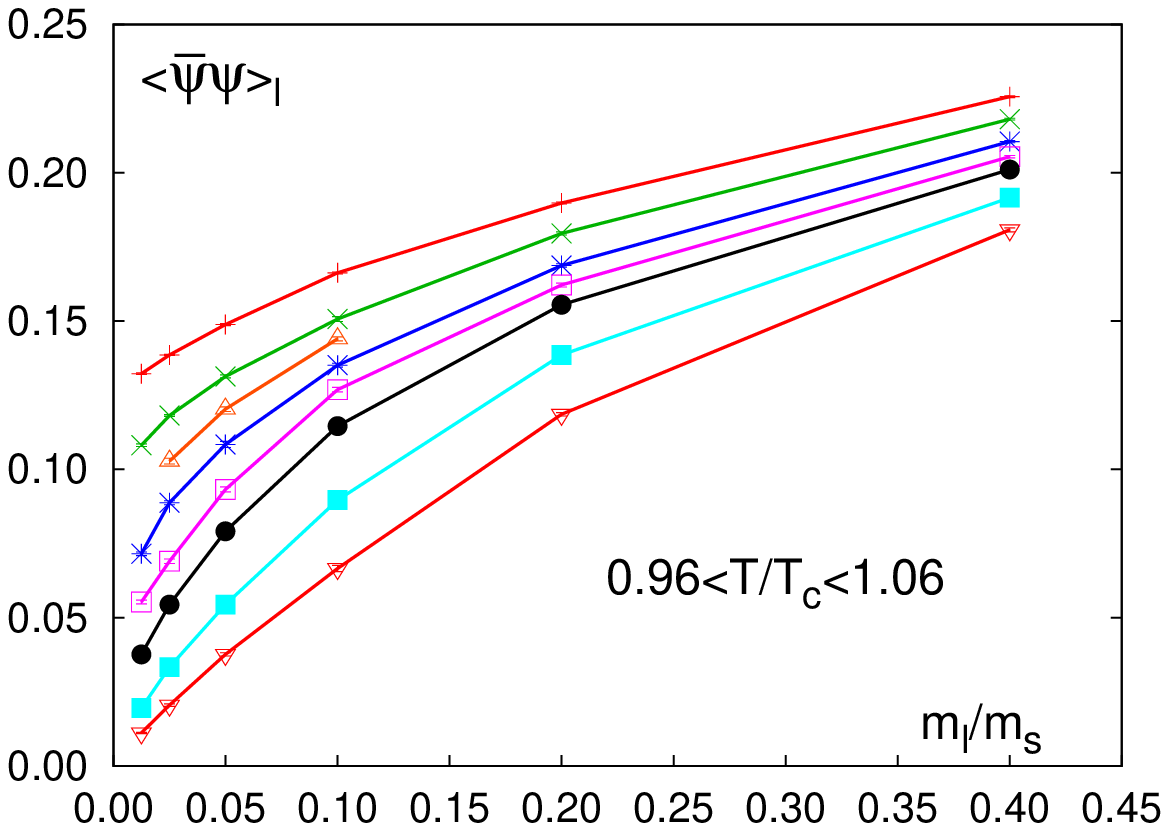,width=9.0cm}\hspace*{-0.4cm}\epsfig{file=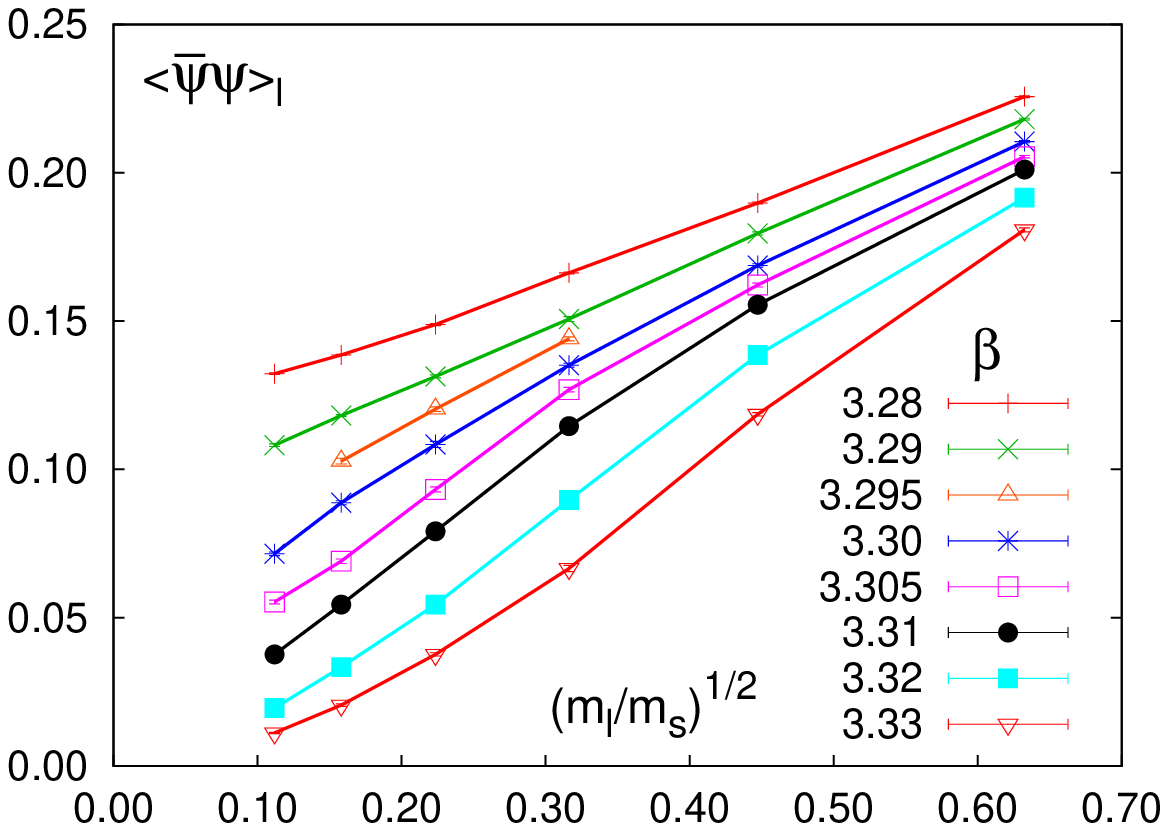,
width=9.0cm}
\end{center}
\caption{\label{fig:pbp_data} The light quark chiral condensate in lattice units versus
the ratio of the light and strange quark masses (left) and its square root (right). The
condensates have been calculated on lattice of size $N_\sigma^3\times 4$ at the
values of the gauge couplings shown in the right hand figure. The spatial lattice size 
for the two lightest quark mass values, $m_l/m_s = 1/80$ and $1/40$, is $N_\sigma=32$, 
for  $m_l/m_s =1/20$ it is $N_\sigma=32$ for $\beta = 3.28$. In all other cases 
($m_l/m_s=1/10,~1/5$ and $2/5$) the spatial lattice size is $N_\sigma = 16$.
}
\end{figure}

A first overview on our data sample is given in the left hand part of 
Fig.~\ref{fig:pbp_data}. This figure shows results for the light quark chiral
condensate in lattice units, calculated
for different values of the gauge coupling and 6 different values of the 
light quark mass. For each parameter set we only show results from the largest 
spatial lattices available. We 
comment on finite volume effects below. We also note that we will conclude
later that the chiral phase transition temperature at $\hm_l=0$ corresponds to
$\beta_c \simeq 3.30$. As is evident from Fig.~\ref{fig:pbp_data} the light quark 
chiral condensate shows a strong
quark mass dependence that is not consistent with a linear dependence on $\hm_l$.
In fact, at low temperatures the dominant quark mass correction seems to be
proportional to $\sqrt{\hm_l}$. This is highlighted in the right hand part of 
Fig.~\ref{fig:pbp_data}. 
Results for the light and strange quark condensates are summarized in
Appendix~\ref{app.data}.

\begin{figure}[t]
\begin{center}
\epsfig{file=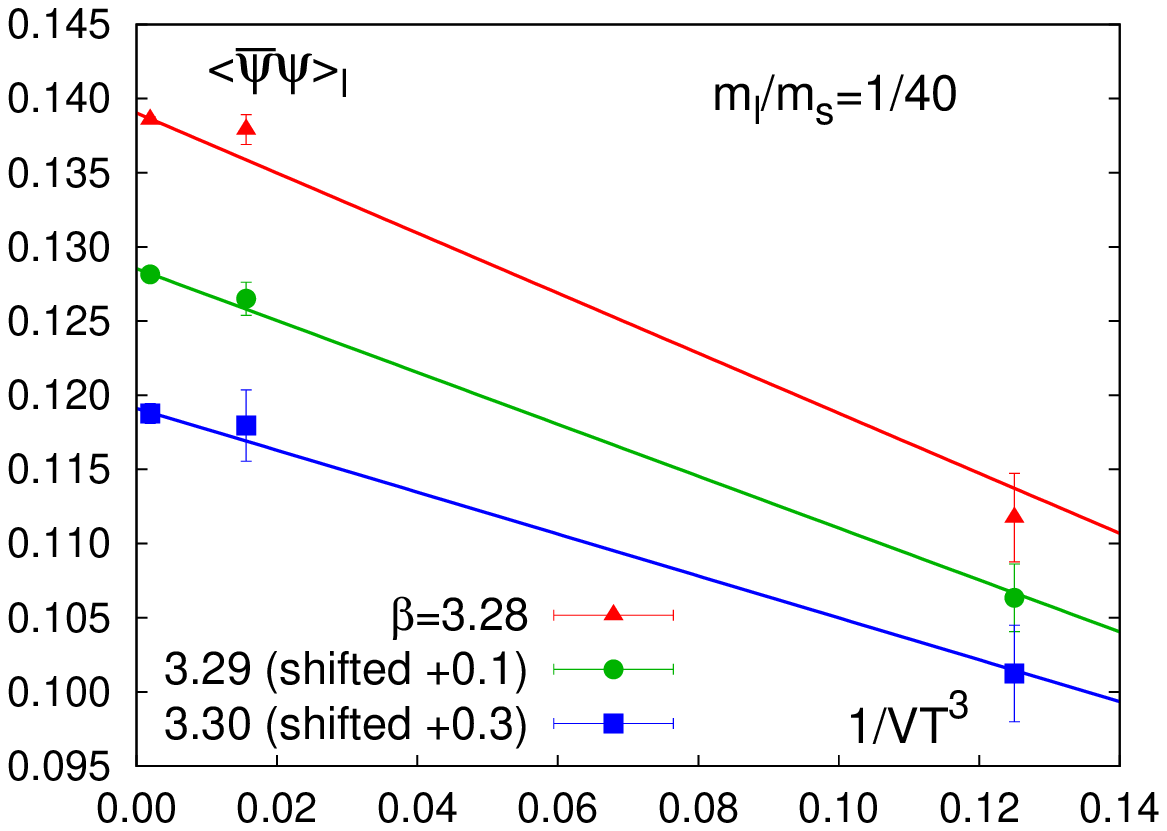,width=9.0cm}\hspace*{-0.4cm}\epsfig{file=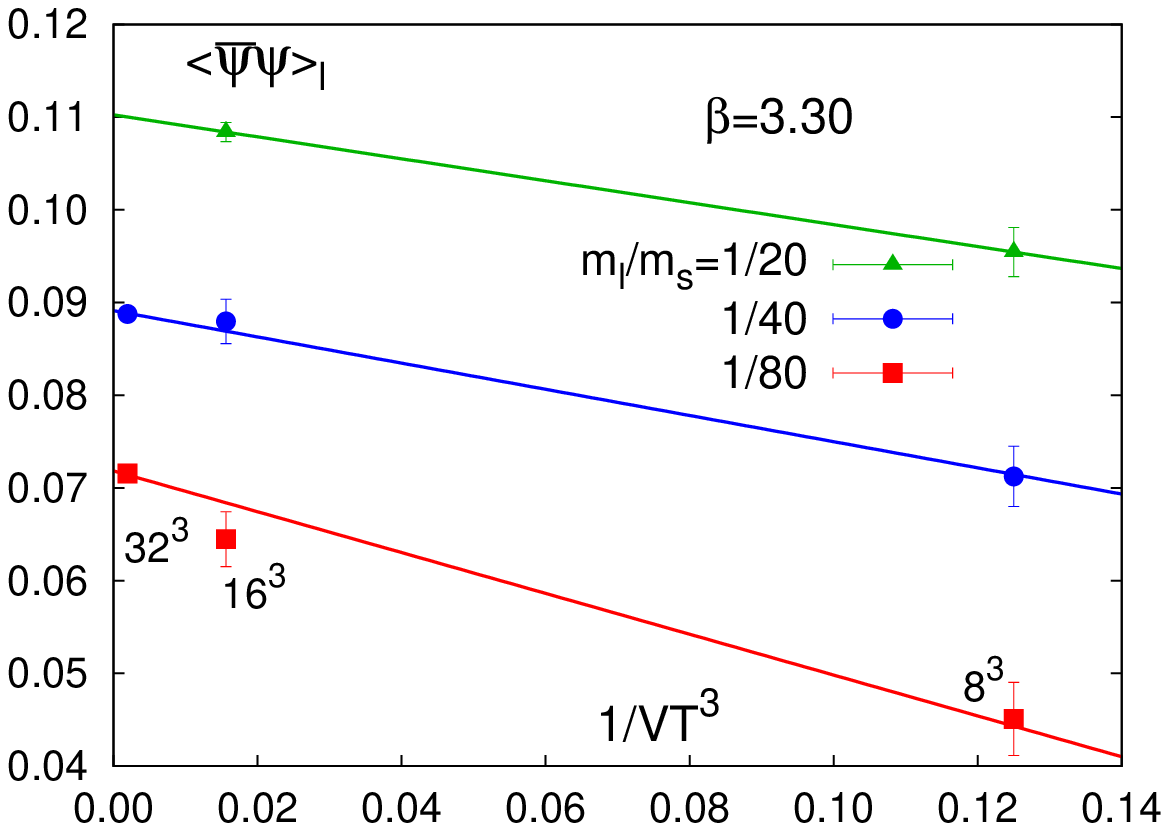,
width=9.0cm}
\end{center}
\caption{\label{fig:pbp_V}
The light quark chiral condensate in lattice units versus the inverse volume, 
$V\equiv N_\sigma^3$, for two values of the gauge coupling in the low temperature
phase, $\beta=3.28,\;3.29$, and close to the transition temperature, $\beta = 3.30$, 
for $m_l/m_s = 1/40$(left).
The right hand figure shows results for the three smallest values of the light quark 
mass at $\beta=3.30$, {\it i.e.} close to the chiral transition temperature.
For these quark mass values the largest lattice size, $N_\sigma=32$,
corresponds to $3 < m_{ps} N_\sigma < 6$. In the left hand figure data for 
$\langle \bar\psi \psi\rangle_l$ have been shifted by a constant as indicated in the 
figure. 
}
\end{figure}

In order to make sure that the drop in $\langle \bar\psi \psi \rangle_l$, 
seen for small values of the quark mass, is not due to a too small spatial 
volume, we analyzed the 
volume dependence of our results by performing calculations on lattices with spatial
extent $N_\sigma = 8,\; 16$ and $32$. 
We show some results from this analysis in Fig.~\ref{fig:pbp_V}. 
As expected, the volume dependence of the 
chiral condensate increases with decreasing value of the quark mass. 
We have, however, no evidence for a strong increase of the volume dependence close to 
the phase transition temperature ($\beta \simeq 3.3$).  
This suggests that also for the smallest quark masses used, our results obtained on lattices 
of spatial size $32^3$ are 
close to the infinite volume limit. We also note that the smallest quark mass value used
on our 
largest lattices corresponds to $m_{ps} N_\sigma \simeq 3$, where $m_{ps}$ denotes the
lightest pseudo-scalar meson mass, {\it i.e.} the Goldstone meson in the staggered
fermion formulation of (2+1)-flavor QCD. 

\subsection{The chiral order parameter}

The chiral condensate, as introduced in Eq.~\ref{condensate},
is an order parameter for the chiral phase transition. At non-vanishing light
quark mass an additive and multiplicative renormalization is needed to
define an order parameter in the continuum limit.
In Ref.~\cite{rbcBIeos} we introduced an order parameter 
where quadratically divergent, additive contributions, 
which are proportional to the quark 
mass, have been removed by subtracting a suitable fraction of the strange
quark chiral condensate from the light quark condensate. We will use this
observable also here\footnote{This does not remove divergencies that are
logarithmic in the cut-off. In the free field limit these divergencies are 
proportional to $(m_l/T)^3$ and are therefore expected to be numerically small}. 
To take into account the anomalous dimensions of the chiral condensate we introduce 
a multiplicative renormalization, using the strange quark mass. 
A similar procedure has been suggested for 2-flavor QCD \cite{Bernard}.
Furthermore, we express this product in units of $T^4$ to make it dimensionless. 
We thus introduce as an order parameter for chiral symmetry restoration in 
(2+1)-flavor QCD
\begin{equation}
M \equiv  \hm_s \left( \langle \bar\psi \psi \rangle_l -
\frac{m_l}{m_s} \langle \bar\psi \psi \rangle_s \right) N_\tau^4 \; .
\label{orderp}
\end{equation}
 
For our scaling analysis at fixed $N_\tau$ a renormalization of the order parameter 
is not at all
necessary. We may as well analyze the scaling behavior of the non-subtracted
chiral condensate introduced in Eq.~\ref{condensate}. To check the 
consistency of our analysis, we will do so and in the following also 
utilize the non-subtracted order parameter
\begin{equation}
M_b = N_\tau^4 \hm_s \langle \bar{\psi}\psi \rangle_l \; ,
\label{Mb}
\end{equation}
where we have introduced the constant multiplicative factor, 
$N_\tau^4 \hm_s$, such that $M_b$ and $M$ agree in the chiral limit.
 
\section{The magnetic equation of state}

Having introduced our numerical results for the light and strange quark chiral 
condensates and the subtracted ($M$) and non-subtracted ($M_b$) order
parameters, we are now ready to discuss critical behavior in the vicinity of 
the transition temperature in terms of the magnetic equation of state.

\subsection{Scaling analysis}

In the vicinity of the chiral phase transition temperature corresponding to 
vanishing light quark masses and for sufficiently small explicit symmetry breaking, 
the order parameter is expected to scale according to Eq.~\ref{order}.
We introduce the reduced temperature
and external field variables $t$ and $h$ as in Eq.~\ref{reduced}. The definition
of $t$ obviously carries over from the spin model context to QCD. For
the relation between the lattice gauge coupling $\beta$ and the 
temperature we exploit the parametrization of
the Sommer scale parameter $r_0$ that has been determined in our 
calculations for the equation of state \cite{rbcBIeos} at $m_l/m_s=1/10$.
This takes into account deviations from asymptotic scaling of the QCD $\beta$-function
for the range of couplings used here. 
We checked that the entire scaling analysis
presented here only for a small range of the lattice cut-off 
is not really sensitive to these corrections and could as well have
been performed by determining $t=(T-T_c)/T_c$ using the asymptotic
2-loop $\beta$-function or other approaches 
followed in earlier studies~\cite{Karsch94,Bernard00,DiGiacomo}.
Likewise this analysis is, of course, independent of any
physical value used to set the absolute scale for $r_0$.

The symmetry breaking external field 
$H$ is proportional to the light quark
mass. Also here we take care of the anomalous scaling dimension of
quark masses and express the light quark mass 
in units of the strange quark mass, 
{\it i.e.} we introduce $H\equiv m_l/m_s$. 
An alternative, yet similar way to deal with the anomalous dimensions
has been suggested in \cite{Bernard}.
 
In the chiral limit, at finite value of the cut-off 
(fixed $N_\tau$) we expect the phase transition in (2+1)-flavor QCD to be either
first order or to belong to the universality class of three dimensional $O(2)$
models. In our current analysis we did not find any indication for a strong volume 
dependence or meta-stabilities in the time evolution of the chiral condensates or
other observables. 
In particular, as is also evident from Fig~\ref{fig:ON_chi} shown in 
Section V, we have no
evidence for a strong increase in the chiral susceptibility as function of volume.
Although we can, at present, not rule out a weak first order 
phase transition at smaller quark masses, we have no indications for that to happen. 
We therefore
will compare our data on the order parameter to the universal $O(2)$ scaling function,
{\it i.e.} the magnetic equation of state introduced in Eq.~\ref{order} with 
critical exponents $\beta$ and $\delta$ given in Table~\ref{tab:parameter}.
We start by determining the three free parameters $t_0$, $h_0$ and the transition
temperature, $T_c$, from fits to the order parameter $M$. For this we use
the three lightest quark mass values, $m_l/m_s \le 1/20$, leaving out 
the lowest temperature value, corresponding to $\beta =3.28$, and the two highest 
temperature values, corresponding to $\beta= 3.32$ and $3.33$. 
For $m_l/m_s=1/80$ and $1/40$ these data are from lattices of size $32^3\times 4$
while the $m_l/m_s=1/20$ data set is taken from calculation on $16^3\times 4$ 
lattices.

\begin{figure}[t]
\begin{center}
\epsfig{file=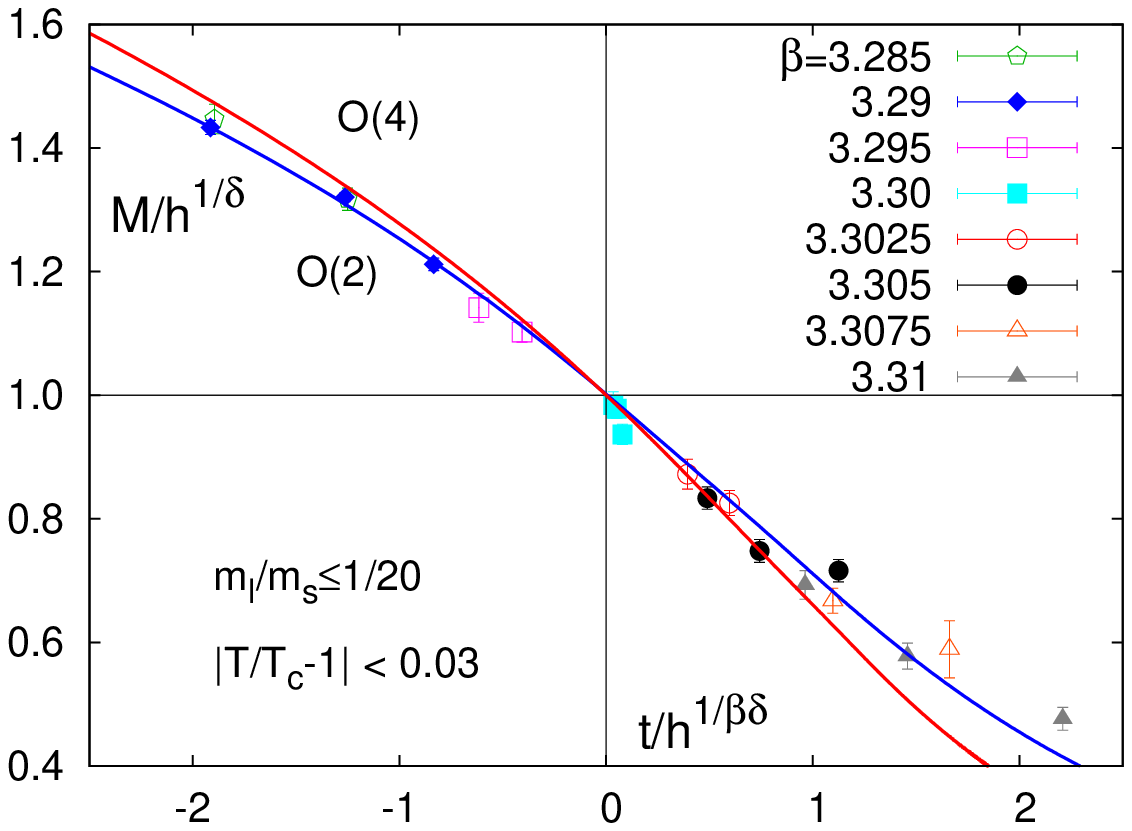,width=9.0cm}\hspace*{-0.4cm}\epsfig{file=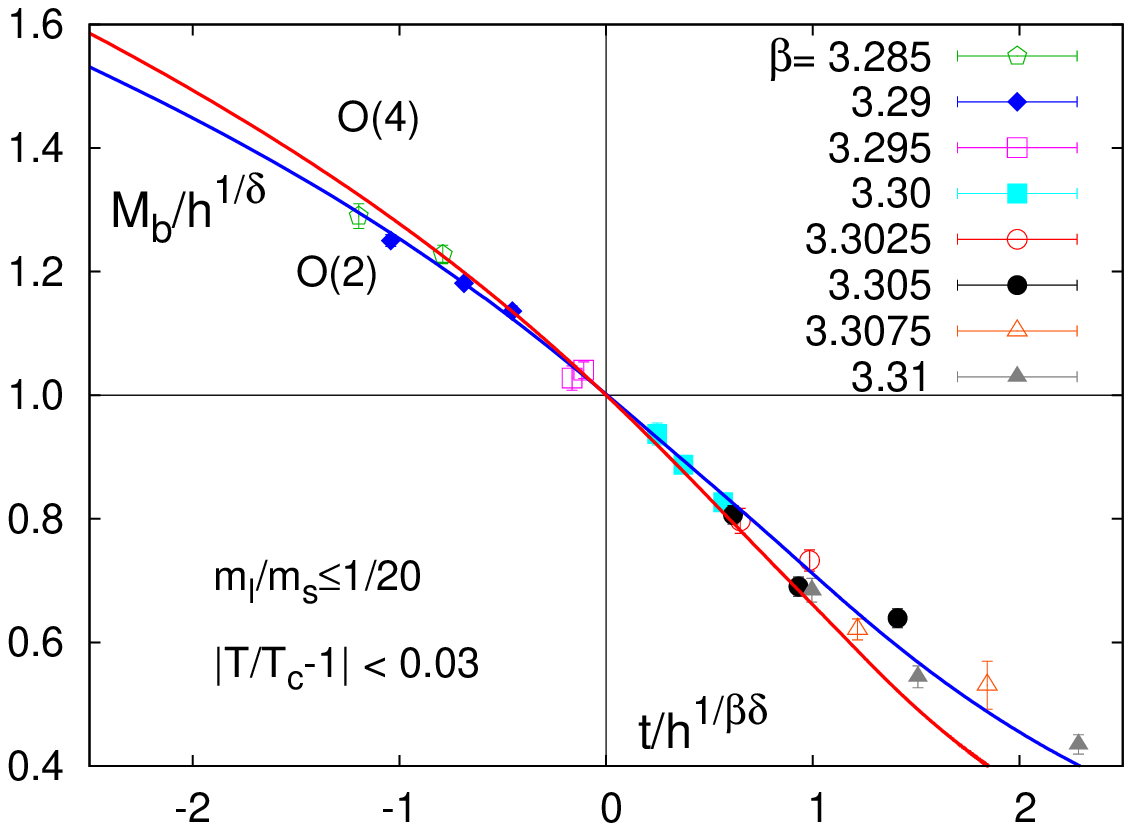,
width=9.0cm}
\end{center}
\caption{\label{fig:magnetic_eos} Fit of the $O(2)$ scaling function to
numerical results for the subtracted order parameter $M$ (left) and the 
non-subtracted light quark condensate $M_b$ (right).
This analysis has been performed for results obtained in
calculations with light quark masses
$m_l/m_s\le 1/20$ and gauge couplings in the interval $\beta\in [3.285,3.31]$.
}
\end{figure}

A posteriori we find that this temperature interval 
corresponds to $0.97 \le T/T_c \le 1.03$. In this small temperature interval and for
the small quark mass regime, which does include the light quark mass value
corresponding to the physical pion mass, we find good agreement between 
the rescaled order parameter and the $O(2)$ scaling function. This is shown in the
left hand part of Fig.~\ref{fig:magnetic_eos}. The fit yields 
$\beta_c =3.300(1)$ for the critical coupling
corresponding to a phase transition temperature\footnote{This 
value is in excellent agreement with our earlier analysis performed with the
p4 action on lattices with temporal extent $N_\tau=4$. 
We stress, however, that this 
transition temperature is not extrapolated to the continuum limit and, in fact, 
within the staggered fermion approach, the chiral extrapolation should be performed
after the continuum extrapolation to recover eventually the anticipated $O(4)$ scaling 
behavior.} $T_c = 195.6(4)$~MeV.
We obtain an equally good agreement with the $O(2)$ scaling curve from an analysis of 
the non-subtracted order parameter $M_b$. This is shown in the right hand part of 
Fig.~\ref{fig:magnetic_eos}. A similar analysis using the $O(4)$ scaling function 
and critical exponents yields scale parameters that are similar to those of the $O(2)$ 
analysis, the largest 
differences occurring for $t_0$ which comes out to be about 20\% larger.

The scaling function $f_G(z)$ is defined in the limit $t\rightarrow 0$, 
$h\rightarrow 0$, keeping $z=t/h^{1/\beta\delta}$ fixed. In this limit
the two order parameters $M$ and $M_b$ coincide. 
From our scaling analysis we therefore should find
identical results for the scale parameters, {\it i.e.}
the critical temperature $T_c$ as well as the 
normalization constants $t_0$, $h_0$, if
this analysis has been performed sufficiently close to the chiral limit.
We have performed the scaling analysis for two different cuts on the 
ratio of the light to strange quark masses, $m_l/m_s\le 1/20$ and 
$m_l/m_s\le 1/40$.
The fit parameters obtained in these two cases 
from an analysis of data for $M$ and $M_b$ 
are summarized in Table~\ref{tab:fit}.
We note that results for $T_c$ and $t_0$ are within errors independent
on the cut on $m_l/m_s$ and the observable used. The scale parameter
$h_0$ is more sensitive on the choice of order parameter. However, there
is a tendency that results for $h_0$ obtained from $M$ and $M_b$ converge
to a common value if the cut on $m_l/m_s$ is reduced.

\begin{table}
\begin{center}
\begin{tabular}{|c|c|c|c|c||c|}
\hline
~&$(m_l/m_s)_{\rm max}$ & $h_0$ & $t_0$ & $T_c$ [MeV] & $z_0$\\
\hline
$M$~  &1/20 & 0.0048(5) & 0.0048(2) & 195.6(4) & 8.5 (7)\\
$M_b$ &1/20 & 0.0022(3) & 0.0037(2) & 194.5(4) & 6.8 (5)\\
\hline
$M$~  &1/40 & 0.0042(6) & 0.0047(2) & 195.3(4) & 8.0 (8)\\
$M_b$ &1/40 & 0.0025(5) & 0.0040(2) & 194.8(4) & 7.0 (6)\\
\hline
\end{tabular}
\end{center}
\caption{\label{tab:fit}
Fit results for the scale parameters $h_0$ and $t_0$ and the 
chiral transition temperature $T_c$ using the $O(2)$ scaling function. 
The last column shows the combination of scale 
parameters $z_0= h_0^{1/\beta\delta} /t_0$.} 
\end{table}

The symmetry breaking field introduced 
in Eq.~\ref{reduced} is given in terms of the ratio of light to strange
quark masses. To compare our result for scaling functions of QCD with other (model)
calculations it may be more convenient to express $H$ in terms of meson masses.
In the present quark mass and gauge coupling range we find the
approximate relation
$H=m_l/m_s \simeq 0.52 \, (m_{ps}/m_K)^2$. 
We therefore may write the scaling variable $z$ as 
\begin{equation}
z =  1.48 \, z_0 \left( \frac{T-T_c}{T_c} \right) / 
\left( \frac{m_{ps}}{m_K} \right)^{2/\beta\delta}  \; .
\label{h0t0}
\end{equation}
As discussed in Section II this allows to determine the scaling behavior of 
the pseudo-critical line determined by the peak in the scaling function
of the chiral susceptibility, $f_\chi$,
\begin{equation}
\frac{T_{p}(m_{ps}) - T_c}{T_c} = 0.68 \frac{z_p}{z_0} 
\left( \frac{m_{ps}}{m_K} \right)^{2/\beta\delta}
\label{pseudo}
\end{equation} 
Using $z_p$ from Table~\ref{tab:parameter} and the values for $z_0$ given in 
Table~\ref{tab:fit} we find $0.68 z_p/z_0 \simeq 0.1-0.2$. 
These values, which are consistent
with earlier determinations of the slope of the pseudo-critical line \cite{Cheng,PeikertTc},
emphasizes the weak dependence of the pseudo-critical temperature on the 
pseudo-scalar meson mass. 

We stress that this analysis has been performed in QCD at 
one non-vanishing lattice spacing, i.e. in the cut-off theory.
The cut-off dependence of the normalization constants, the scale 
invariant ratio $z_0$ and the subtle
continuum limit need to be studied in the future.
We emphasize again, however, that the 
above combination of normalization constants for the scaling variables 
is an invariant of QCD and depends, in the continuum limit, only on the strange
quark mass value.

\subsection{Scaling violations}

\begin{figure}[t]
\begin{center}
\epsfig{file=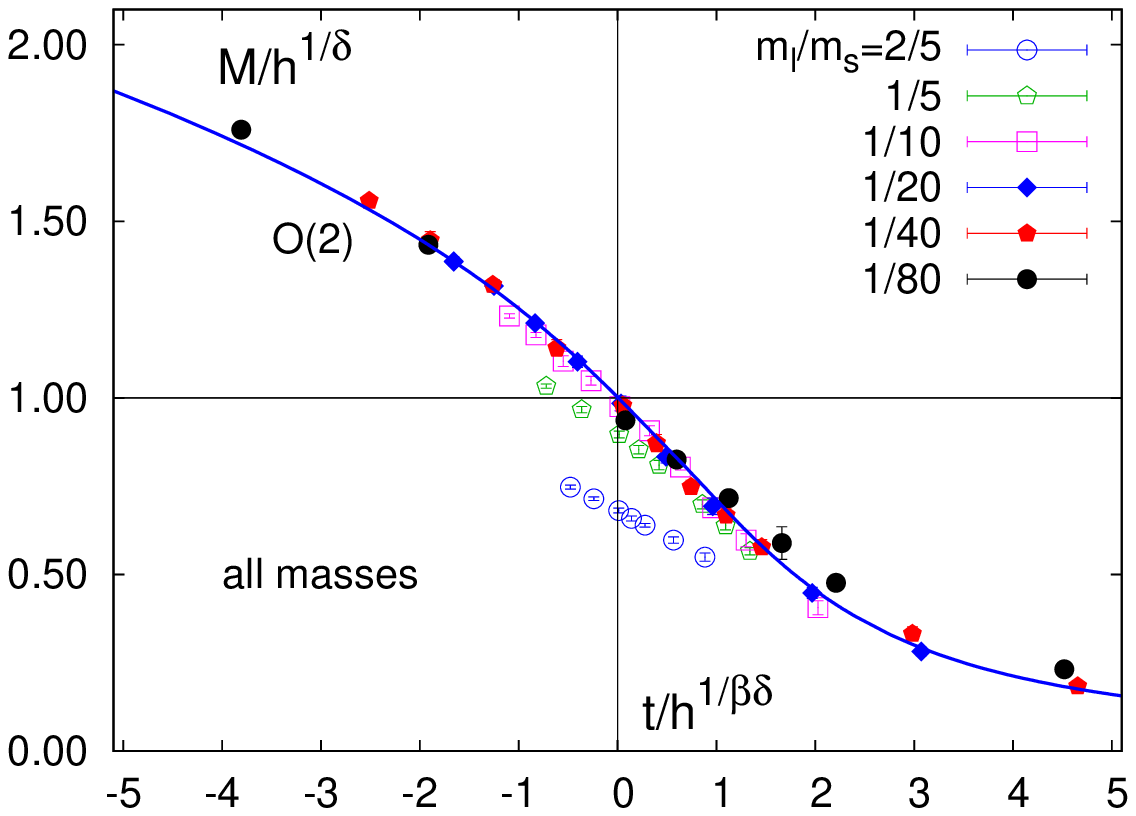,width=9.0cm}\hspace*{-0.4cm}\epsfig{file=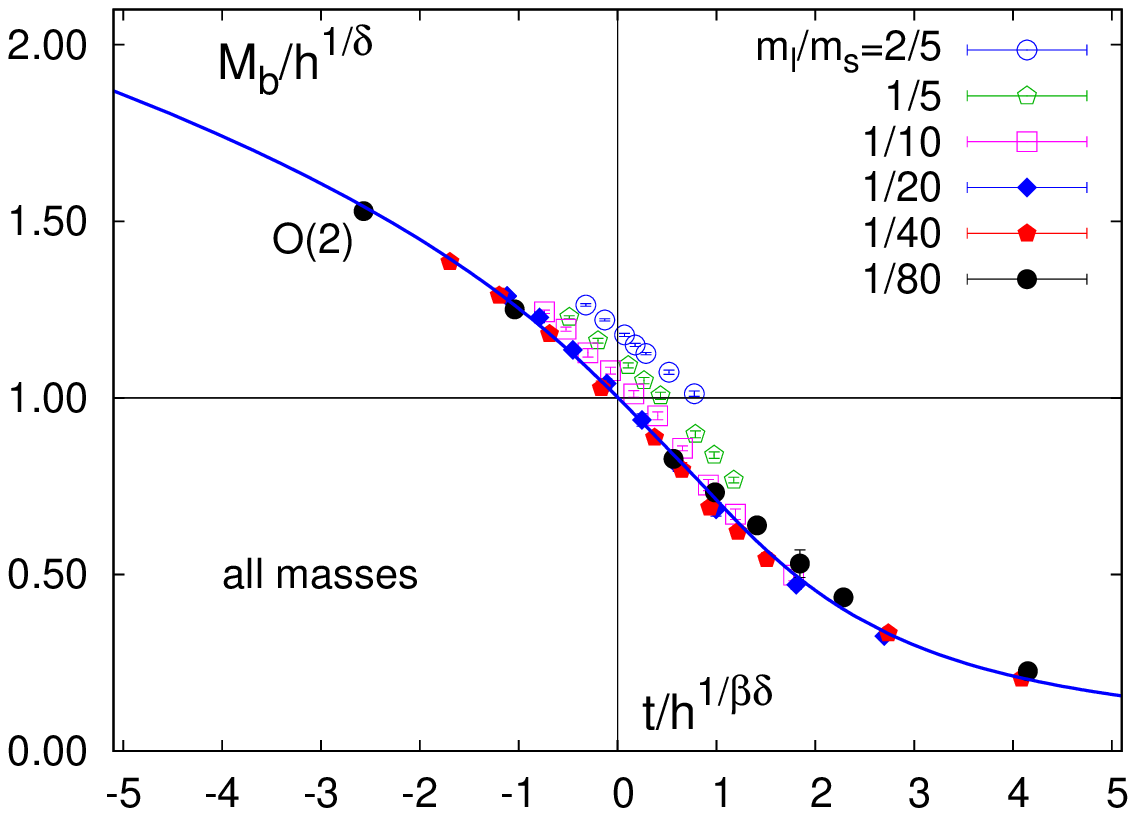,
width=9.0cm}
\end{center}
\caption{\label{fig:magnetic_eos_all} 
The order parameters $M$ (left) and $M_b$ (right)
for all quark mass 
values, $m_l/m_s\le 0.4$, and all values of the gauge coupling, 
$\beta\in [3.28,3.33]$, used in this study. 
The scaling variables $t$ and $h$ used to compare with the $O(2)$ scaling function
are taken from the fit to the light quark mass results shown in 
Fig.~\protect\ref{fig:magnetic_eos}.
}
\end{figure}

The scaling behavior observed for the chiral order parameters analyzed in the 
previous section is, of course, expected to hold exactly only in the limit
$t\rightarrow 0$ and $h\rightarrow 0$, keeping the ratio $z=t/h^{1/\beta\delta}$
fixed. At non-zero values of $t$ and $h$ we expect to observe scaling violations
that may arise from sub-leading corrections to the scaling function as well as 
from the regular part of the QCD partition function. 
These corrections also depend on the definition of the order
parameter. In particular, the two order parameters, $M$ and $M_b$, introduced 
here differ in the treatment of contributions that are linear in the light
quark mass. In our analysis of the order parameter, performed in a larger temperature 
and quark mass interval, we clearly see 
these differences and their role in contributing 
to violations of scaling. This is shown in Fig.~\ref{fig:magnetic_eos_all}.
Most prominent are effects arising from a too large quark mass value. These effects
show up in the scaling plot as deviations from the scaling function in the region
of small $z$, {\it i.e.} for large quark masses at fixed $t$. They lead to the 
sizeable displacement of results obtained for too heavy quarks from the scaling 
curve.  Effects that arise because the temperatures chosen are too far away from the 
critical point, $t=0$, are typically not that drastic in our data sample. We fitted 
the scaling violations to an ansatz
\begin{equation}
M(t,h) = h^{1/\delta} f_G(t/h^{1/\beta\delta}) + a_t t h +  
b_1 h + b_3 h^3 +b_5 h^5\; .
\label{violations}
\end{equation}
We also considered including a term quadratic in the reduced temperature ($\sim t^2 h$).
This correction, however, turned out to vanish within the errors of our fits.

\begin{figure}[t]
\begin{center}
\epsfig{file=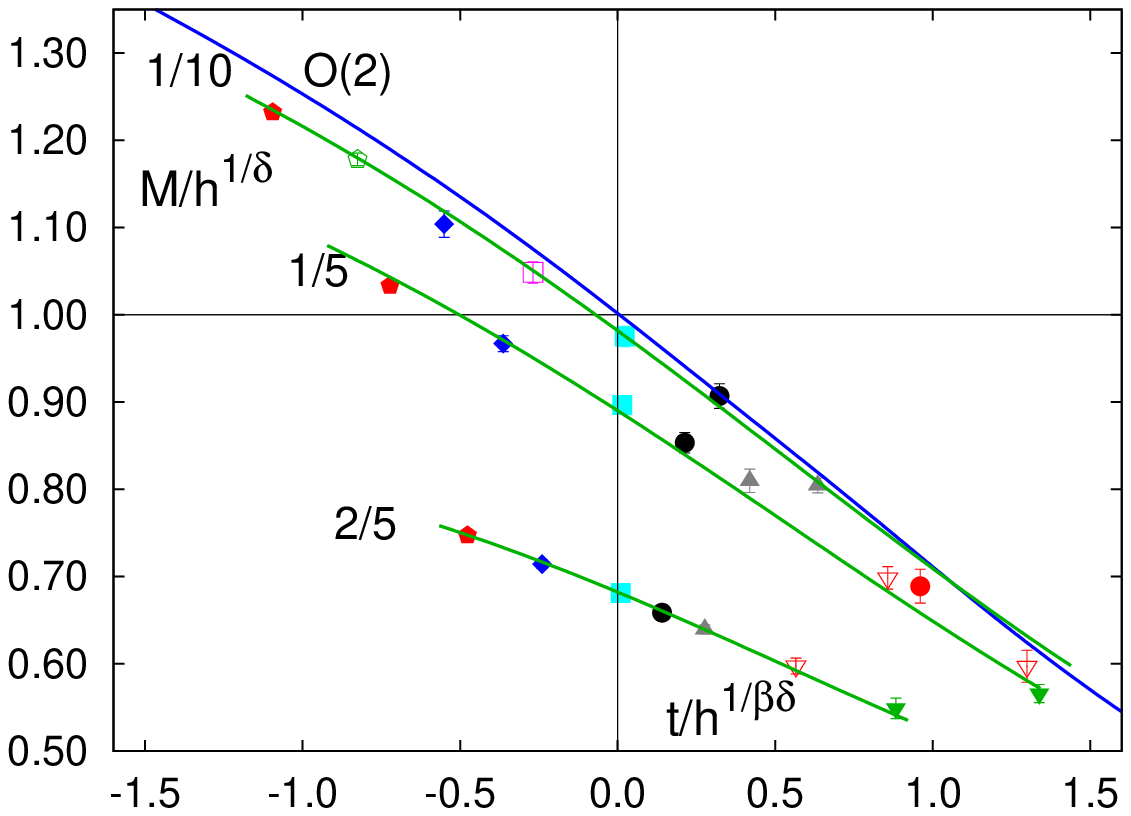,width=9.0cm}\hspace*{-0.4cm}\epsfig{file=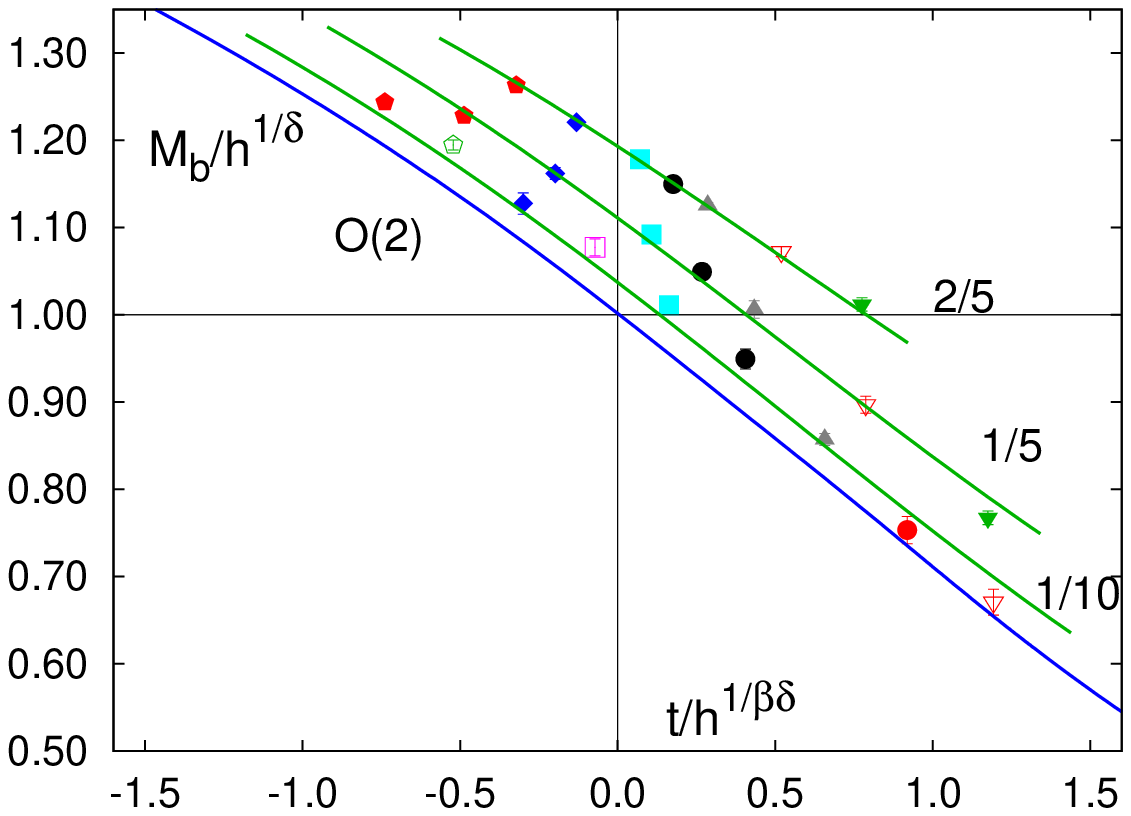,width=9.0cm}
\end{center}
\caption{\label{fig:violations} The $O(2)$ magnetic 
equation of state compared to results 
for the subtracted order parameter $M$ (left) and the non-subtracted chiral 
condensate, $M_b$ for light quark masses $m_l/m_s \ge 1/10$. 
Curves show fits to data at fixed $m_l/m_s$ using the ansatz for 
scaling violations given in Eq.~\ref{violations}. 
Same symbols correspond to same values of the gauge coupling. 
}
\end{figure}

The fits of both order parameters performed with the ansatz given in 
Eq.~\ref{violations} are shown in Fig.~\ref{fig:violations}.
As expected, we find that corrections linear in $m_l/m_s$ are eliminated
in $M$. The corresponding fit parameter $b_1$ is zero within errors and 
we therefore have fixed it to be zero in the fit shown in 
Fig.~\ref{fig:violations} (left). For the non-subtracted order parameter $M_b$
this term gives the dominant finite quark mass corrections. Here we find
$b_1 = 0.0013(3)$. 

\subsection{Scaling of the chiral condensate}

\begin{figure}[t]
\begin{center}
\epsfig{file=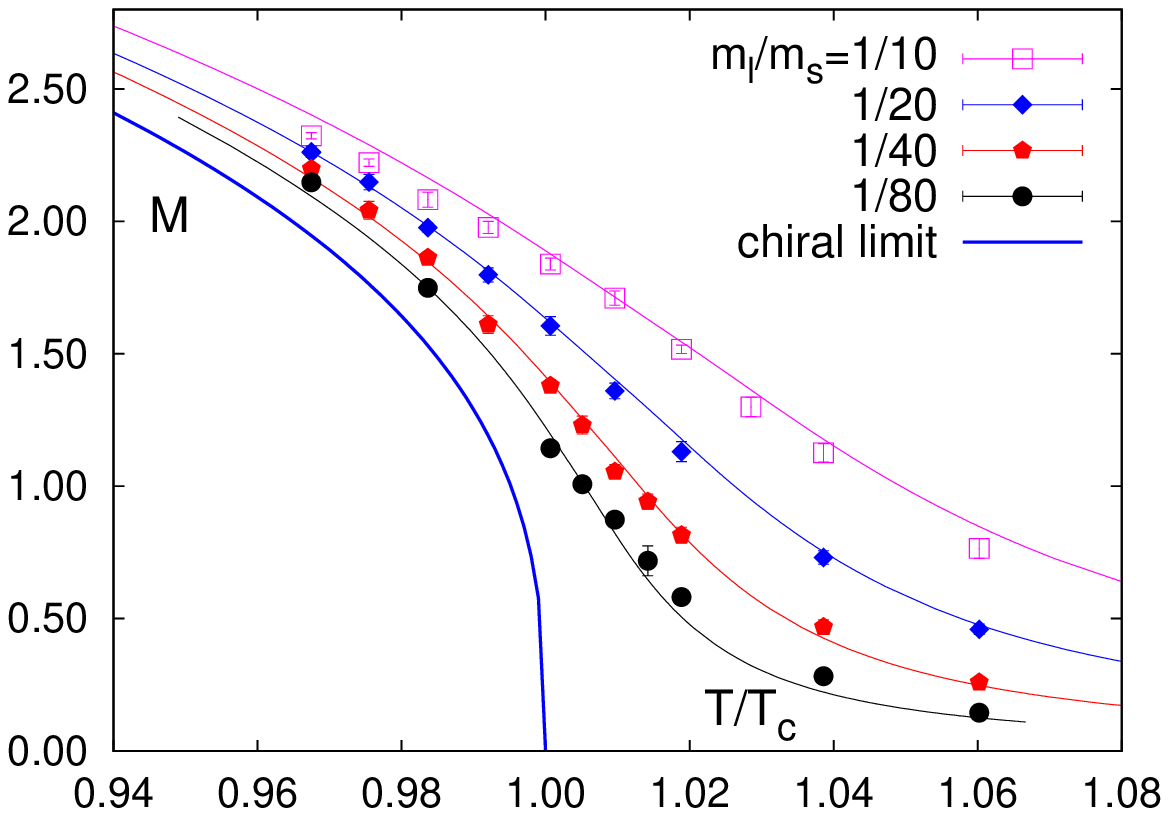,width=9.0cm}\hspace*{-0.2cm}\epsfig{file=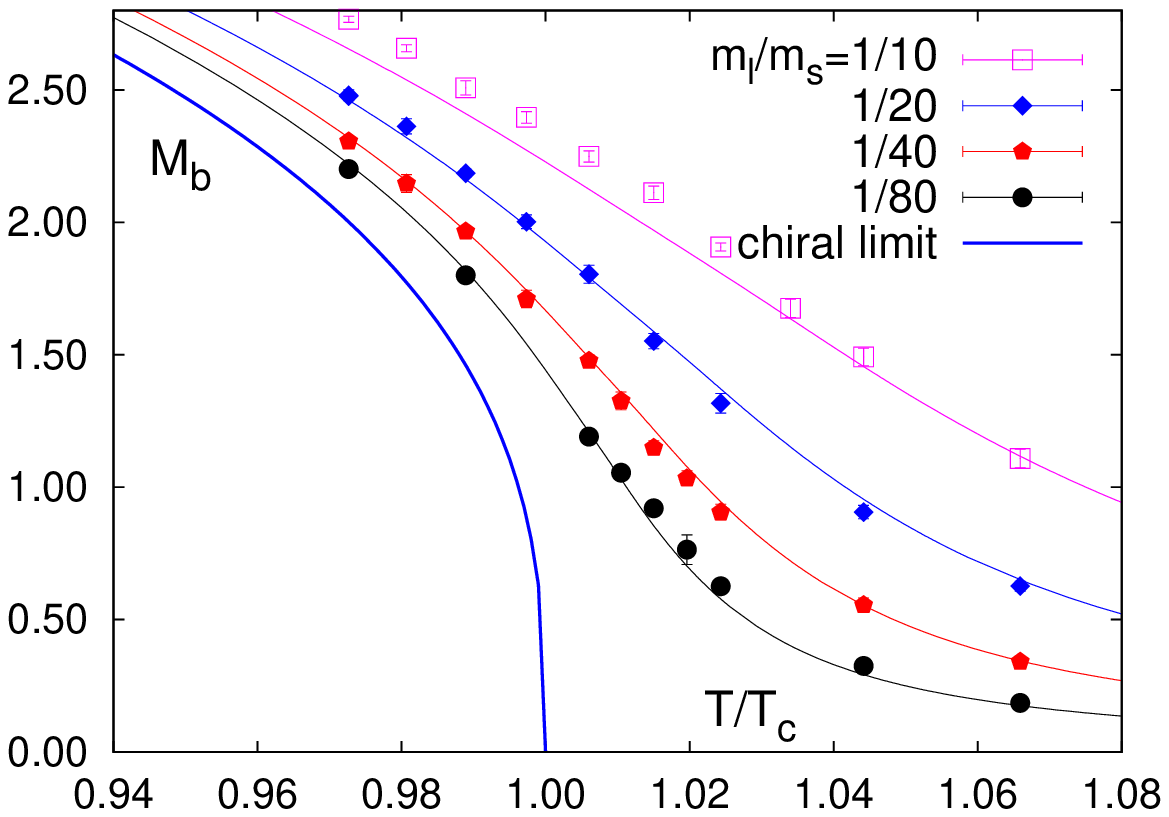,width=9.0cm}
\end{center}
\caption{\label{fig:condensate} The subtracted chiral order 
parameter, $M$, defined in Eq.~\ref{orderp}, 
compared to the fit result for the magnetic equation of state (left). 
The right hand figure
shows results for the un-subtracted, but normalized chiral condensate 
$M_b$ defined in Eq.~\ref{Mb}. 
}
\end{figure}

We have seen in the previous section that order parameters constructed from the chiral 
condensate are well described
by the magnetic equation of state for small enough values of the light quark masses, 
$m_l/m_s\lsim 1/20$. We want to underscore this point here by displaying
the order parameters not in their scaling form, but as a function of temperature 
in units of the transition
temperature determined in the previous section. This is shown in  
Fig.~\ref{fig:condensate}.  The curves drawn
in this figure are taken from the scaling fits
to the subtracted and non-subtracted order parameters shown in
Fig.~\ref{fig:magnetic_eos}. They had been obtained from the numerical 
results for $M$ (left) and $M_b$ (right) in the range $m_l/m_s\le 1/20$ 
and $T/T_c=1 \pm0.03$. 

\subsection{Comparison with earlier calculations in 2-flavor QCD}

As mentioned in the Introduction, there have been earlier attempts 
to compare the quark 
mass and temperature dependence of the chiral order parameter with $O(N)$ scaling 
functions on lattices with temporal extent $N_\tau=4$ 
\cite{Bernard00,DiGiacomo,Mendes}. 
These calculations had been performed for 2-flavor QCD using unimproved gauge and 
staggered fermion actions. In Ref.~\cite{Bernard00} calculations with 
three quark mass 
values had been performed, $\hm = 0.008,\; 0.0125$ and $0.025$. The last two 
masses are similar to the two mass values used in Ref.~\cite{DiGiacomo}, {\it i.e.}
$\hm = 0.01335$ and $0.0267$. In fact, results for chiral condensates obtained in
these two calculations are in good agreement with each other. This also is true
for calculations performed in \cite{KL} where $\hm =0.02$ has been used.
All these calculations have been performed at values of the gauge coupling in the 
vicinity of the cross-over at the corresponding quark mass values. 
They therefore mostly explored the region of $z>0$.

\begin{figure}[t]
\begin{center}
\epsfig{file=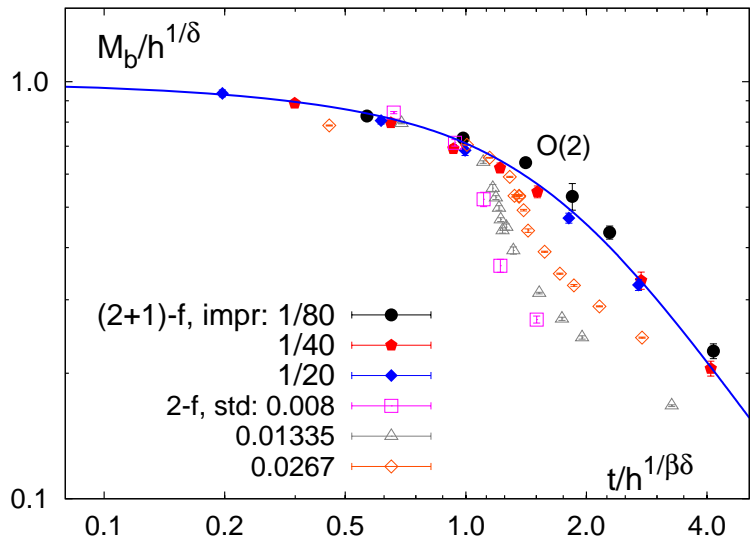,width=11.0cm}
\end{center}
\caption{\label{fig:loglog} Scaling plot for the chiral condensate 
calculated with an improved staggered action in (2+1)-flavor QCD  (this work) and 
the standard staggered action in 2-flavor QCD \cite{Bernard00,DiGiacomo}. The results 
are shown in a log-log plot. For the (2+1)-flavor data set labels indicate the 
ratio $m_l/m_s$, in the 2--flavor case we give the bare quark masses $\hm$. Data
for the lightest quark mass are from \cite{Bernard00}. Data for the other two quark 
mass values are from \cite{DiGiacomo}. For further discussion see text.
}
\end{figure}

In Fig.~\ref{fig:loglog} we compare results for the chiral condensate obtained 
in 2-flavor calculations with unimproved gauge and fermion actions with
our results obtained in (2+1)-flavor QCD with ${\cal O}(a^2)$ improved gauge
and fermion actions. In this figure we use a log-log plot as has been done 
also in Ref.~\cite{Bernard00}. 
In the 2-flavor case the symmetry breaking field has usually been chosen
as $H=\hm N_\tau$ while for the reduced temperature variable we used
$(T-T_c)/T_c \equiv R(\beta_c)/R(\beta)-1$, with $\beta_c=5.2435$ as 
estimate for the critical value of the gauge coupling in the chiral limit 
\cite{DiGiacomo} 
and $R(\beta$) denoting the 2-loop $\beta$-function for 2-flavor
QCD. In the log-log plot differences in the scale 
parameters $h_0$ and $z_0$ correspond to shifts in vertical and horizontal 
directions, respectively. We made no effort to optimize the choice of these scale 
parameters for the 2-flavor data set. In Fig.~\ref{fig:loglog} we have
positioned the data such that the crossover region roughly corresponds
to the location of the maximum in the $O(2)$ scaling function $f_\chi(z_p)$,
with $z_p=1.56$ (see also \cite{Mendes}); this required the choice $z_0\simeq 12$.

When comparing results obtained with standard and improved gauge actions 
the difference in the shape of the data sets clearly is the most striking feature.
Apparently the 
(2+1)-flavor data set is in good agreement with $O(N)$ scaling while the results
obtained with the standard staggered action deviate strongly. Furthermore, there is
no tendency for better agreement with decreasing quark mass. Closer to the
continuum limit, {\it i.e.} for larger $N_\tau$, results obtained with the 
unimproved actions have a similar shape \cite{Bernard00} but seem to get somewhat 
closer to the $O(N)$ scaling curve.

To compare the quark mass values used in the 2-flavor QCD calculations with those
of the present (2+1)-flavor study we note that for standard staggered fermions 
at gauge couplings close to the cross-over 
a bare quark mass $\hm=0.025$ corresponds to a pseudo-scalar Goldstone mass 
$m_{ps}\simeq 350$~MeV \cite{Blum}.  
The lightest quark mass used in these calculations, 
$\hm =0.008$, therefore corresponds to $m_{ps}\simeq 200$~MeV, which is similar to the
pseudo-scalar mass obtained in (2+1)-flavor QCD calculations for the light to strange 
quark mass ratio $m_l/m_s = 1/10$. At this value for the light quark mass we observe
only mild violations of scaling in calculations with the improved gauge and fermion
actions.

The number of flavors as well as the quark masses are different in the
data sets compared in Fig.~\ref{fig:loglog}.  Nevertheless, it seems unlikely that 
this is the origin of the observed differences. It appears more probable
that cut-off effects 
in calculations with unimproved gauge and fermion actions cause the differences. 

\section{Susceptibilities}

In Eq.~\ref{chiM} we introduced the susceptibility $\chi_M$ as the derivative
of the order parameter $M$ with respect to the symmetry breaking field. Its 
temperature and quark mass dependence is controlled by the scaling function 
$f_\chi (z)$ defined in Eq.~\ref{fchi}, which is shown in the right hand part 
of Fig.~\ref{fig:scaling} for $O(2)$ and $O(4)$. Similarly we can, of course, introduce
the susceptibility $\chi_{Mb}$ as a derivative of the order parameter $M_b$ with respect
to $H$.
Taking derivatives with respect to $h\equiv m_l/(m_s h_0)$, rather than with respect to
the ratio of quark masses, $ m_l/m_s$,
obviously requires knowledge of the scale parameter $h_0$ which we have determined
in the previous section.

In the analysis of QCD thermodynamics on the lattice it is more customary to
calculate light ($\chi_m^l$) and strange ($\chi_m^s$) quark chiral 
susceptibilities, which are defined as derivatives of the corresponding chiral 
condensates with respect to $m_l/T$ and $m_s/T$, respectively
\begin{eqnarray}
\chi_m^q/T^2 &=& N_\tau^3 \frac{{\rm d}\langle \bar\psi \psi \rangle_q}
{{\rm d} (m_q/T)} 
\; ,\; q=l,~s \\
\label{chiralsus}
\end{eqnarray}
To construct the susceptibility $\chi_M$ we will also need to take into account 
a mixed chiral susceptibility,
\begin{eqnarray}
\chi_m^{ls}/T^2 &=& N_\tau^3 \frac{{\rm d}\langle \bar\psi \psi \rangle_s}
{{\rm d} (m_l/T)} 
\label{mixed}
\end{eqnarray}
The susceptibility of the subtracted order parameter, $M$, is then obtained as
\begin{eqnarray}
\chi_M &=& \frac{\partial M}{\partial h}  \nonumber \\
&=& h_0 N_\tau^2 \hm_s^2  
\left( \frac{\chi_m^l}{T^2} - \frac{N_\tau^2}{\hm_s} \langle \bar\psi \psi \rangle_s -
\frac{m_l}{m_s} \frac{\chi_m^{ls}}{T^2} \right) 
\nonumber \\
&=& \chi_{Mb} - h_0 N_\tau^4 \hm_s \langle \bar\psi \psi \rangle_s -
h_0 N_\tau^2 \hm_s^2 \frac{m_l}{m_s} \frac{\chi_m^{ls}}{T^2} \; ,
\label{chiMr}
\end{eqnarray}
where in the last equality we introduced the susceptibility $\chi_{Mb}$, of the 
non-subtracted order parameter $M_b$. 
With this we can construct the scaling function for the chiral susceptibility,
\begin{eqnarray}
f_\chi(z) &=& \chi_M h_0/ h^{1/\delta-1}
\equiv h_0^{1/\delta} \left( \frac{m_l}{m_s} \right)^{1-1/\delta} \chi_M \; ,
\label{chi_scaling}
\end{eqnarray}
from $\chi_M$ and similarly also from $\chi_{Mb}$.

\begin{figure}[t]
\begin{center}
\epsfig{file=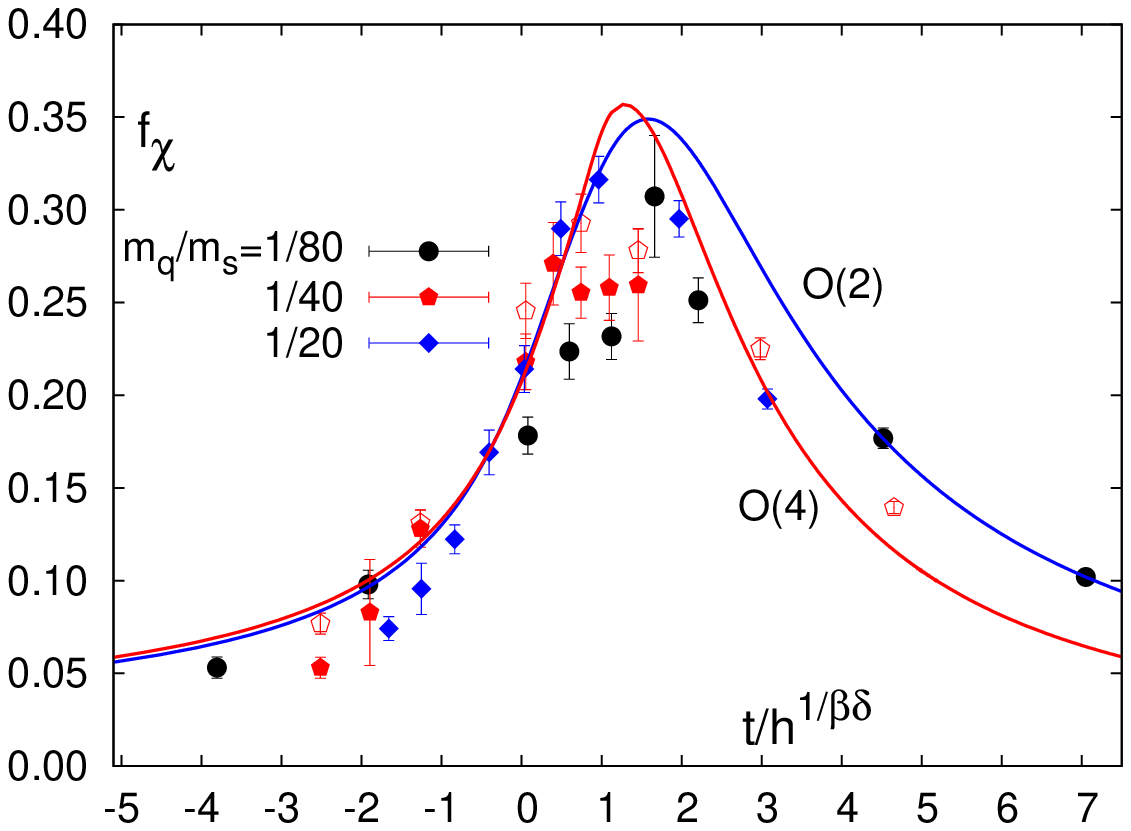,width=9.0cm}\hspace*{-0.4cm}\epsfig{file=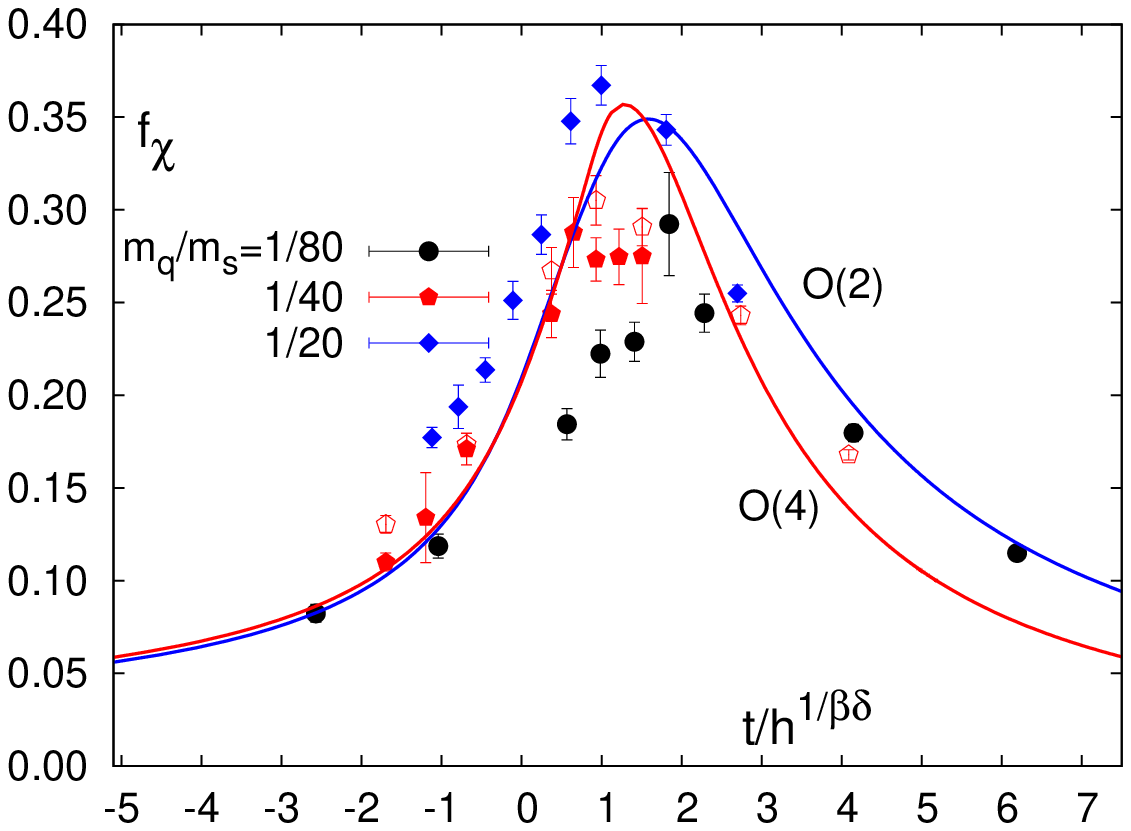,width=9.0cm}
\end{center}
\caption{\label{fig:ON_chi} 
Susceptibilities constructed from the subtracted order parameter $M$ (left) 
and the non-subtracted light quark chiral condensate $M_b$ (right).
The data give results from calculations in (2+1)-flavor
QCD on lattices with temporal extent $N_\tau=4$ and light quark mass values 
$m_l/m_s \le 1/20$ in the interval $\beta \in [3.28,3.33]$. For the $m_l/m_s=1/40$
data sample we show results for two different spatial lattice sizes. Filled
symbols correspond to $N_\sigma =32$ and open symbols are for $N_\sigma = 16$.
}
\end{figure}

The scaling functions $f_\chi$, constructed from  either $\chi_M$ or $\chi_{Mb}$,
differ by terms that vanish in the chiral limit at $T_c$. These
terms therefore characterize once more systematic differences that arise in the
construction of scaling functions due to the presence of regular terms that
vanish, once the appropriate scaling limits are taken. We show scaling functions
constructed from both order parameters in Fig.~\ref{fig:ON_chi}. We stress that
all parameters ($T_c$, $t_0$ and $h_0$) have been determined in our analysis
of the order parameters themselves. No fits are therefore involved in the
comparison of the $O(N)$ scaling functions with the numerical results for
susceptibilities shown in this figure. 

It is obvious from Fig.~~\ref{fig:ON_chi} that violations of scaling are
significantly larger for susceptibilities than for the order parameters.
Susceptibilities extracted from $M$ and $M_b$ still differ for $m_l/m_s=1/20$
but start to become compatible within errors for $m_l/m_s\le 1/40$. 
For $m_l/m_s=1/40$ we also show results from calculations on two different
lattice size, $32^3\times 4$ and $16^3\times 4$. If any, the volume dependence
of susceptibilities is small at this value of the quark mass.

There is yet another difference between order parameter susceptibilities 
derived in QCD, where the order parameter is a composed operator constructed
from fermionic fields, and $O(N)$ spin models, where the order parameter
is the expectation value of a scalar boson field. The order parameter susceptibilities
in QCD receive two contributions, usually called the disconnected and connected
part of the susceptibility,
\begin{equation}
\chi_m^l \equiv  2 \chi_l^{\rm dis} + \chi_l^{\rm con} \; ,
\label{chi_dis_con}
\end{equation}
with
\begin{eqnarray}
\chi_{\rm dis} &=& {1 \over 16 N_{\sigma}^3 N_{\tau}} \left\{
\langle\bigl( {\rm Tr} D_l^{-1}\bigr)^2  \rangle -
\langle {\rm Tr} D_l^{-1}\rangle^2 \right\}
\label{chi_dis} \; , \\
\chi_{\rm con} &=&  {1 \over 4} \sum_x \langle \,D_l^{-1}(x,0) D_l^{-1}(0,x) \,\rangle \; .
\label{chi_con}
\end{eqnarray}
While the first term, the disconnected part of the light quark susceptibility,
describes fluctuations of the light quark condensate and has a direct analogy
in the fluctuations of the order parameter in an $O(N)$ spin model, the second
term ($\chi_{\rm con}$) arises from the explicit quark mass dependence of the
order parameter, the chiral condensate. 
The connected part is an integral over
the (quark-line connected) correlation function of the 
(iso-vector) scalar operator, $\bar{\psi}\psi$. 
The integral has a rather subtle quark mass dependence. 
Since $\delta > 2$, however, $\chi_{\rm con}/h^{1/\delta -1}$ will 
vanish in the chiral limit.  In this limit
the connected part of the susceptibility will 
therefore not contribute to the scaling function $f_\chi$ 
which in turn will entirely be determined through the
disconnected part of the light quark susceptibility.

At non-vanishing values of the light quark mass, however, 
the non-vanishing connected part of the chiral susceptibility is responsible 
for additional scaling violations in $f_\chi$.
In fact, the scaling violations due to the connected part
are distinctively different between $O(2)$ and $O(4)$ symmetric
theories \cite{Smilga}. It is only in the latter 
that fluctuations of Goldstone modes do 
not contribute to $\chi_{\rm con}$. In the case of $O(2)$ symmetric models
$\chi_{\rm con}$ is expected to diverge proportional to $1/\sqrt{h}$ just
like the total order parameter susceptibilities will do. 

In the staggered formulation of QCD with 2 light quark flavors
the lack of $O(4)$ symmetry in the 
Lagrangian is due to explicit symmetry breaking terms (taste violations)
that disappear only in the continuum limit.
Corresponding to the $O(2)$ spin models,
at finite values of the cut-off
the divergence of $\chi_{\rm con} \sim 1/\sqrt{\hm_l}$ in the chiral limit 
can thus be understood in terms of taste violating contributions to the 
scalar correlation function \cite{Prelovsek}.
We will discuss these subtle aspects of susceptibilities of the order parameter,
the influence of taste violating terms in the staggered action on scaling
properties of these susceptibilities and the resulting cut-off dependence
of $f_\chi$ in more detail in a forthcoming publication \cite{staggered}.

\section{Conclusions}

We have performed a new analysis of scaling properties of the light quark
chiral condensate in $(2+1)$-flavor QCD. We  found that 
at fixed non-zero lattice spacing the chiral condensate calculated with 
improved staggered fermions shows scaling behavior in the chiral limit that is 
consistent with $O(2)$ scaling. 

Through the analysis of scaling properties with quark masses that are smaller 
than the physical light quark masses we could fix the normalization constants
$t_0$ and $h_0$ in the scaling variables $t$ and $h$.  This allowed us to quantify 
scaling violations for non-zero values of the quark masses in the vicinity of the 
phase transition temperature.  These scaling violations turned out to be
small in the magnetic equation of state already 
for physical values of the quark mass. 

On the basis of studying just the magnetic equation of state,
we gave arguments that it will remain difficult 
to rule out $O(4)$ scaling without extraordinary precision of
numerical lattice data. However, a distinction between $O(2)$ and
$O(4)$ scaling might become possible through an accurate
analysis of susceptibilities of the order parameter. 
At present, we still find
significant deviations from scaling for the chiral susceptibility.
This will be discussed in more detail in a forthcoming 
publication \cite{staggered}.

A determination of $t_0$ and $h_0$ also fixes the scale parameter 
$z_0 = h_0^{1/\beta\delta}/t_0$, which  controls the quark mass 
dependence of the pseudo-critical line determined from the peak in the 
chiral susceptibility. In our present analysis this parameter,
which uniquely characterizes non-universal aspects of critical
behavior in QCD, has only been determined at one value of the 
lattice cut-off. Calculations 
at smaller lattice spacings, together with good
control over scaling violations
induced at non-vanishing quark masses will be needed to extract $z_0$ in
the $O(4)$ symmetric continuum limit.

The good scaling properties found here in calculations with ${\cal O}(a^2)$
improved gauge and fermion actions are
in contrast to earlier calculations that had 
been performed with unimproved staggered fermion and gauge actions. We argued
that the observed differences are due to cut-off effects.

In our analysis we have assumed that the strange quark mass in $(2+1)$-flavor
QCD is large enough to avoid a first order phase transition in the light quark
chiral limit. Although the good scaling properties of the chiral order parameters
and the absence of a strong volume dependence in the 
light quark susceptibilities support this assumption, we clearly cannot 
exclude a first order transition to occur at still lighter quark masses. 
Consistent with limits given on the location
of a first order transition in 3-flavor
QCD \cite{our3f,fodor}, however,
our current analysis rules out such a 
transition for pseudo-scalar masses $m_{ps} \ge 75$~MeV.

\section*{Acknowledgments}
\label{ackn}
This work has been supported in part by contracts DE-AC02-98CH10886
with the U.S. Department of Energy,
the BMBF under grant 06BI401, the Gesellschaft
f\"ur Schwerionenforschung under grant BILAER, the Extreme Matter Institute
under grant HA216/EMMI and the Deutsche
Forschungsgemeinschaft under grant GRK 881. Numerical simulations have
been performed on the BlueGene/L at the New York Center for Computational 
Sciences (NYCCS) which is supported by the U.S. Department
of Energy and by the State of New York as well as the QCDOC computer of USQCD. 
We thank J. Engels for discussions and for providing us with his programs to calculate
the $O(N)$ scaling functions.

%\newpage
\appendix

\section{Scaling functions for three-dimensional \boldmath $O(2)$ and $O(4)$ models}
\label{app.scaling}

In this appendix we summarize the scaling functions for models in the
three-dimensional $O(2)$ and $O(4)$ universality classes. These interpolating functions
have been taken from Ref.~\cite{Engels2001} and \cite{Engels2003}. We note that the
original parameters for the $O(4)$ model published in the tables of Ref.~\cite{Engels2001}
had been updated in \cite{Engels2003}. Moreover, interpolating curves had been
constructed only for the scaling function $f_G$. Applying these interpolations
also to $f_\chi$ required slight adjustments of the interpolation parameters
($y_0$, $p$).

In Eq.~\ref{order} we expressed the dependence of the order parameter $M$ on the
scaling variables $t$ and $h$ in terms of a scaling function, $f_G$.
Following the discussion given in \cite{Engels2001} we introduce the variables $x$
and $y$,
\begin{equation}
y  =  f_G^{-\delta} \;, \quad
x   =  (t/h^{1/\beta\delta}) f_G^{-1/\beta}\;.
\label{xy}
\end{equation}
Obviously $y\ge 0$. For small and large values of $y$ the asymptotic
forms that relate $x$ to $y$ are known. For small $y$ we have
\begin{equation}
x_s(y) = -1+ ({\widetilde c_1} \,+\, {\widetilde d_3})\,y \,+\,
             {\widetilde c_2}\,y^{1/2} \,+\,
             {\widetilde d_2}\,y^{3/2} \;,
\label{xs}
\end{equation}
and for large values of $y$ one finds
\begin{equation}
x_l(y)  =   a\, y^{1/\gamma} + b\,  y^{(1-2\beta)/\gamma}~.
\label{xl}
\end{equation}
One can smoothly interpolate between these two relations \cite{Engels2001}
using the ansatz

\begin{equation}
x(y) \;=\; x_s(y)\,\frac{y_0^p}{y_0^p + y^p} \,+\,
           x_l(y)\,\frac{y^p}{y_0^p + y^p}~.
\label{xinterpolate}
\end{equation}
This ansatz has been used to obtain the scaling functions shown
in Fig.~\ref{fig:scaling} for $-0.5\lsim \,z\,\lsim 2.0$. For $|z|$ outside this
interval the asymptotic expressions $x_l (y)$ and $x_s(y)$ have been used.
The constant $\widetilde c_2$ and the critical exponents $\beta,\, \gamma,\, \delta$ 
are given in Table~\ref{tab:parameter}, the other
parameters needed for this interpolation are collected
in Table~\ref{tab:inter}.
\begin{table}[htib]
\begin{center}
  \begin{tabular}{|c|cc||cc||cc|}
    \hline
~& ${\widetilde c_1} \,+\, {\widetilde d_3}$ &
$ {\widetilde d_2}$ & $a$ & $b$ &  $y_0$ & $p$~ \\ \hline \hline
$O(2)$ & 0.352(30)& 0.056 & 1.260(3)& -1.163(20) &   ~2.5 & 
 3~  \\ \hline
$O(4)$ & 0.359(10)&-0.025(10) & 1.071(4)& -0.866(38) & 5.0 & 3~ \\ \hline
  \end{tabular}
\end{center}
\caption{Parameters of the fits to the scaling functions for $O(2)$
and $O(4)$.}
\label{tab:inter}
\end{table}

\section{Chiral condensates from calculations on lattices with temporal extent \boldmath $N_\tau=4$}
\label{app.data}

In this appendix we present data of our calculations performed with the p4 staggered
fermion action on lattices with temporal extent $N_\tau =4$ and spatial extent 
$N_\sigma =8$, $16$ and $32$. All calculations have been performed with 2 light
quarks and a strange quark of mass $\hm_s =0.065$. The improved gauge and fermion
actions used for these calculations have been described in detail in Ref.~\cite{Peikert}.
The tables give results of calculations performed at different values of the gauge 
coupling ($\beta$). 
Results for light ($l$) and strange ($s$) quark condensates as well as the 
disconnected and connected contributions to the corresponding susceptibilities
are normalized to a single flavor. The last column gives the number of 
trajectories generated for each parameter set. 

In Table~\ref{tab:littlet} we give for some selected values of the gauge coupling
$\beta$ the conversion to a reduced temperature scale.

\newpage

\begin{table}[b]
\begin{center}
\begin{tabular}{|@{\extracolsep{3mm}}c|ccc|ccc|c|}      
\hline  
\hspace{8mm}$\beta$\hspace{8mm} & $\expval{\bar{\psi}\psi}_l$ & $\chi_l^{\rm con}$ & $\chi_l^{\rm dis}$ & $\expval{\bar{\psi}\psi}_s$ & $\chi_s^{\rm con}$ & $\chi_s^{\rm dis}$ & \# traj.\\  
\hline  
\multicolumn{8}{|c|}{$N_\sigma^3\times N_\tau = 32^3\times 4$, $m_l a=0.0008125$} \\    
\hline  
3.2800 & 0.1322(3) & 3.90(16) & 2.43(24) &      0.2575(1) & 1.412(1) & 0.47(4) & 18730~\\
3.2900 & 0.1082(4) & 4.73(12) & 3.95(34) &      0.2454(1) & 1.477(1) & 0.63(5) & 20070~\\
3.3000 & 0.0715(4) & 7.44(11) & 6.10(44) &      0.2294(1) & 1.575(1) & 0.63(4) & 18830~\\
3.3025 & 0.0633(6) & 8.46(13) & 7.61(67) &      0.2261(2) & 1.593(1) & 0.77(7) & 15810~\\
3.3050 & 0.0553(5) & 9.44(12) & 7.46(55) &      0.2228(2) & 1.613(1) & 0.78(7) & 17460~\\
3.3075 & 0.0459(15) & 11.30(35) & 9.91(1.47) &  0.2190(5) & 1.634(3) & 1.04(17) & ~4530~\\
3.3100 & 0.0376(5) & 12.31(14) & 6.85(53) &     0.2156(1) & 1.656(1) & 0.70(6) & 15850~\\
3.3200 & 0.0195(4) & 12.98(17) & 3.07(23) &     0.2054(2) & 1.706(2) & 0.57(4) & 10380~\\
3.3300 & 0.0111(2) & 10.48(11) & 0.88(10) &     0.1968(1) & 1.745(1) & 0.40(4) & ~6850~\\
\hline  
\multicolumn{8}{|c|}{$N_\sigma^3\times N_\tau = 32^3\times 4$, $m_l a=0.0016250$} \\    
\hline  
3.2800 & 0.1386(2) & 3.15(5) & 1.81(14) &       0.2590(1) & 1.400(1) & 0.42(3) & 21080~\\
3.2850 & 0.1290(8) & 3.28(20) & 2.48(73) &      0.2536(3) & 1.433(3) & 0.51(13) & ~3400~\\
3.2900 & 0.1181(3) & 3.80(6) & 3.36(26) &       0.2479(1) & 1.460(1) & 0.62(5) & 20940~\\
3.2950 & 0.1009(12) & 4.41(20) & 2.28(49) &     0.2396(5) & 1.505(4) & 0.41(6) & ~1900~\\
3.3000 & 0.0888(4) & 5.10(6) & 4.95(39) &       0.2336(1) & 1.543(1) & 0.74(6) & 20550~\\
3.3025 & 0.0797(5) & 5.63(8) & 6.04(57) &       0.2295(2) & 1.568(2) & 0.89(9) & 16870~\\
3.3050 & 0.0690(4) & 6.47(6) & 5.17(35) &       0.2249(1) & 1.599(1) & 0.69(5) & 21280~\\
3.3075 & 0.0621(7) & 6.94(9) & 4.98(45) &       0.2217(2) & 1.616(2) & 0.63(6) & ~6570~\\
3.3100 & 0.0545(7) & 7.45(9) & 4.74(78) &       0.2183(3) & 1.633(2) & 0.72(14) & ~7370~\\
\hline  
\multicolumn{8}{|c|}{$N_\sigma^3\times N_\tau = 16^3\times 4$, $m_l a=0.0016250$} \\    
\hline  
3.2800 & 0.1380(7) & 3.77(12) & 2.12(13) &      0.2593(3) & 1.401(2) & 0.44(3) & 21180~\\
3.2900 & 0.1165(7) & 4.47(12) & 3.10(17) &      0.2475(3) & 1.462(2) & 0.55(3) & 27440~\\
3.3000 & 0.0880(10) & 5.74(11) & 5.35(38) &     0.2337(4) & 1.541(3) & 0.76(6) & 40000~\\
3.3050 & 0.0688(10) & 7.11(10) & 5.83(40) &     0.2252(4) & 1.599(3) & 0.82(7) & 42000~\\
3.3100 & 0.0533(11) & 8.14(13) & 4.88(30) &     0.2182(4) & 1.636(3) & 0.63(4) & 24910~\\
3.3200 & 0.0334(7) & 9.05(8) & 2.95(14) &       0.2075(3) & 1.692(2) & 0.51(3) & 25050~\\
3.3300 & 0.0202(4) & 8.16(9) & 1.09(6) &        0.1975(3) & 1.740(2) & 0.39(3) & 14870~\\
\hline  
\multicolumn{8}{|c|}{$N_\sigma^3\times N_\tau = 16^3\times 4$, $m_l a=0.0032500$} \\    
\hline  
3.2800 & 0.1487(4) & 2.84(8) & 1.73(8) &        0.2615(2) & 1.374(3) & 0.44(2) & 40360~\\
3.2850 & 0.1421(9) & 2.93(10) & 1.98(20) &      0.2575(5) & 1.405(5) & 0.49(5) & 13260~\\
3.2900 & 0.1308(5) & 3.20(3) & 2.20(11) &       0.2510(2) & 1.433(2) & 0.50(2) & 45080~\\
3.2950 & 0.1204(9) & 3.56(7) & 2.69(17) &       0.2454(4) & 1.469(4) & 0.60(4) & 19110~\\
3.3000 & 0.1083(6) & 3.88(3) & 3.16(18) &       0.2388(3) & 1.504(2) & 0.64(3) & 41050~\\
3.3050 & 0.0933(7) & 4.43(5) & 3.97(21) &       0.2312(3) & 1.551(3) & 0.76(4) & 39960~\\
3.3100 & 0.0792(7) & 4.82(5) & 4.12(18) &       0.2239(3) & 1.586(3) & 0.77(3) & 42890~\\
3.3200 & 0.0542(6) & 5.88(4) & 3.16(14) &       0.2107(2) & 1.668(2) & 0.66(3) & 44490~\\
3.3300 & 0.0336(3) & 6.25(2) & 1.41(8) &        0.1984(2) & 1.734(1) & 0.46(2) & 39320~\\
\hline
\multicolumn{8}{c}{}\\
\hline
\multicolumn{8}{|c|}{$N_\sigma^3\times N_\tau = 32^3\times 4$, $m_l a=0.0032500$} \\
\hline
3.2800 & 0.1488(2) & - & 1.61(13) &       0.2615(1) & - & 0.46(3) & 20000~\\
\hline
\multicolumn{8}{|c|}{$N_\sigma^3\times N_\tau = 16^3\times 4$, $m_l a=0.0008125$} \\
\hline
3.3000 & 0.0627(25) & - & 7.64(94) &      0.2279(9) & - & 0.71(10) & ~6690~\\
\hline
\multicolumn{8}{|c|}{$N_\sigma^3\times N_\tau = 8^3\times 4$, $m_l a=0.0008125$} \\
\hline
3.3000 & 0.0452(21) & - & 4.68(39) &      0.2335(13) & - & 0.70(5) & 25830~\\
\hline
\multicolumn{8}{|c|}{$N_\sigma^3\times N_\tau = 8^3\times 4$, $m_l a=0.0016250$} \\
\hline
3.2800 & 0.1141(29) & - & 5.26(25) &      0.2584(12) & - & 0.51(4) & 38280~\\
3.2900 & 0.0963(21) & - & 5.24(17) &      0.2485(9) & - & 0.61(3) & 40660~\\
3.3000 & 0.0753(35) & - & 5.32(34) &      0.2380(16) & - & 0.75(6) & 40100~\\
\hline
\multicolumn{8}{|c|}{$N_\sigma^3\times N_\tau = 8^3\times 4$, $m_l a=0.0032500$} \\
\hline
3.3000 & 0.0954(21) & - & 3.49(13) &      0.2372(10) & - & 0.63(3) & 30000~\\
\hline
\end{tabular} 
\end{center}
\caption{\label{tab:low} Light and strange quark condensates
($\langle \bar{\psi}\psi\rangle_{l,s}$) for $m_l/m_s\le 1/20$ and the corresponding 
disconnected and connected parts of the chiral
susceptibilities. The last column gives the number of trajectories of half unit length
generated for each parameter set. 
}
\end{table}

\begin{table}[b]
\begin{center}
\begin{tabular}{|@{\extracolsep{3mm}}c|ccc|ccc|c|}      
\hline  
\hspace{8mm}$\beta$\hspace{8mm} & $\expval{\bar{\psi}\psi}_l$ & $\chi_l^{\rm con}$ & $\chi_l^{\rm dis}$ & $\expval{\bar{\psi}\psi}_s$ & $\chi_s^{\rm con}$ & $\chi_s^{\rm dis}$ & \# traj.\\  
\hline  
\multicolumn{8}{|c|}{$N_\sigma^3\times N_\tau = 16^3\times 4$, $m_l a=0.0065000$} \\    
\hline  
3.2800 & 0.1660(4) & 2.37(1) & 1.11(6) &        0.2661(2) & 1.352(1) & 0.37(2) & 27550~\\
3.2850 & 0.1595(5) & 2.50(2) & 1.14(8) &        0.2619(2) & 1.371(2) & 0.38(2) & 19150~\\
3.2900 & 0.1507(4) & 2.60(1) & 1.37(7) &        0.2562(2) & 1.402(2) & 0.43(2) & 30160~\\
3.2950 & 0.1439(5) & 2.74(2) & 1.77(12) &       0.2519(3) & 1.424(2) & 0.53(3) & 24880~\\
3.3000 & 0.1351(5) & 2.85(1) & 1.94(12) &       0.2464(2) & 1.452(2) & 0.57(3) & 36100~\\
3.3050 & 0.1269(5) & 3.01(4) & 2.23(12) &       0.2414(3) & 1.471(5) & 0.63(3) & 40230~\\
3.3100 & 0.1146(6) & 3.22(4) & 2.74(15) &       0.2343(3) & 1.515(5) & 0.74(4) & 40440~\\
3.3150 & 0.1007(6) & 3.53(3) & 3.09(15) &       0.2262(3) & 1.555(4) & 0.81(4) & 45570~\\
3.3200 & 0.0895(6) & 3.80(2) & 2.53(15) &       0.2197(3) & 1.597(3) & 0.65(4) & 33360~\\
3.3300 & 0.0666(8) & 4.29(2) & 2.10(14) &       0.2061(4) & 1.678(3) & 0.62(4) & 18060~\\
\hline  
\multicolumn{8}{|c|}{$N_\sigma^3\times N_\tau = 16^3\times 4$, $m_l a=0.0130000$} \\    
\hline  
3.2800 & 0.1899(3) & 1.98(0) & 0.59(3) &        0.2722(2) & 1.317(1) & 0.28(1) & 20050~\\
3.2900 & 0.1795(4) & 2.07(0) & 0.79(4) &        0.2645(2) & 1.350(1) & 0.34(1) & 21040~\\
3.3000 & 0.1687(5) & 2.18(1) & 1.06(7) &        0.2565(3) & 1.387(3) & 0.45(3) & 18880~\\
3.3050 & 0.1621(4) & 2.27(1) & 1.03(6) &        0.2518(2) & 1.413(4) & 0.43(2) & 32170~\\
3.3100 & 0.1555(6) & 2.31(1) & 1.05(9) &        0.2470(4) & 1.429(2) & 0.44(3) & 16580~\\
3.3200 & 0.1383(6) & 2.55(2) & 1.67(10) &       0.2354(3) & 1.492(4) & 0.64(4) & 28740~\\
3.3250 & 0.1295(4) & 2.72(3) & 1.81(9) &        0.2295(2) & 1.530(13) & 0.68(3) & 54840~\\
3.3300 & 0.1186(5) & 2.81(1) & 1.92(9) &        0.2222(3) & 1.566(2) & 0.72(3) & 50000~\\
\hline  
\multicolumn{8}{|c|}{$N_\sigma^3\times N_\tau = 16^3\times 4$, $m_l a=0.0260000$} \\    
\hline  
3.2800 & 0.2256(2) & 1.62(0) & 0.39(2) &        0.2812(1) & 1.267(1) & 0.24(1) & 20310~\\
3.2900 & 0.2180(2) & 1.66(0) & 0.42(2) &        0.2747(2) & 1.290(1) & 0.25(1) & 23950~\\
3.3000 & 0.2105(3) & 1.71(0) & 0.45(2) &        0.2683(2) & 1.314(2) & 0.27(1) & 15330~\\
3.3050 & 0.2054(3) & 1.73(0) & 0.44(3) &        0.2642(2) & 1.331(1) & 0.27(1) & 22550~\\
3.3100 & 0.2010(3) & 1.76(1) & 0.57(4) &        0.2605(2) & 1.346(4) & 0.34(2) & 20170~\\
3.3200 & 0.1916(4) & 1.82(0) & 0.67(4) &        0.2528(3) & 1.378(1) & 0.39(2) & 20030~\\
3.3300 & 0.1807(4) & 1.90(0) & 0.74(6) &        0.2440(3) & 1.417(2) & 0.41(3) & 23380~\\
\hline  
\end{tabular}
\end{center}
\caption{\label{tab:high} Same as Table~\protect\ref{tab:low} but for the heavier
quark masses, $m_l/m_s \ge 1/10$. 
}
\end{table}

\begin{table}
\begin{center}
\begin{tabular}{|c|rrrrrr|}     
\hline 
$\beta$     & 3.2800 & 3.2900 & 3.3000 & 3.3100 & 3.3200 & 3.3300 \\
\hline 
$(T-T_c)/T_c$ & -0.0332 & -0.0170 & 0.0000 & 0.0181 & 0.0379 & 0.0595\\
\hline 
\end{tabular}
\end{center}
\caption{\label{tab:littlet}
Relation between $\beta$ and $(T-T_c)/T_c$ for $\beta_c=3.3000$.
A shift in $\beta_c$ of 0.001 corresponds to a shift in $(T-T_c)/T_c$ of about
0.0017.
}
\end{table}

%\newpage

\end{document}